\documentclass[twocolumn,tighten]{aastex62}

\usepackage[english]{babel}
\usepackage{boldline}
\usepackage[intlimits]{amsmath}
\usepackage{bm}
\usepackage{gensymb}
\usepackage[caption=false]{subfig}
\usepackage{txfonts}
\usepackage{dsfont}
\usepackage{verbatim}
\usepackage{etoolbox}
\usepackage{float}
\usepackage{nicefrac}
\usepackage{booktabs}
\usepackage{varioref}				
\usepackage{hyperref}				
\DeclareRobustCommand{\DUTCH}[3]{#2}    

\graphicspath{{figures/}}

\usepackage{array}
\renewcommand{\arraystretch}{1.2}

\labelformat{chapter}{Chap.~#1}
\labelformat{section}{Sec.~#1}
\labelformat{appendix}{App.~#1}
\labelformat{subsection}{Sec.~#1}
\labelformat{subsubsection}{Sec.~#1}
\labelformat{figure}{Fig.~#1}
\labelformat{subfigure}{Fig.~\thefigure #1}
\labelformat{table}{Tab.~#1}
\labelformat{equation}{Eq.~(#1)}

\usepackage[procnames]{listings}
\usepackage{textcomp}
\usepackage{numprint}

\definecolor{text}{HTML}{000000}
\definecolor{keyword}{HTML}{0000FF}
\definecolor{builtin}{HTML}{900090}
\definecolor{definition}{HTML}{000000} 
\definecolor{comment}{HTML}{ADADAD} 
\definecolor{string}{HTML}{00AA00}
\definecolor{number}{HTML}{800000}
\definecolor{instance}{HTML}{924900} 
\definecolor{linenumber}{HTML}{ADADAD} 

\definecolor{green}{rgb}{0,1,00}
\definecolor{lightgreen}{rgb}{0,0.5,0}
\definecolor{red}{rgb}{1,0,0}
\definecolor{lightred}{rgb}{0.5,0,0}
\newcounter{inCounter}[section]

\makeatletter
\newif\iffirstchar\firstchartrue
\newif\ifstartedbyadigit

\newcommand\ProcessLetter
{%
	\ifnum\lst@mode=\lst@Pmode%
		\iffirstchar%
				\global\startedbyadigitfalse%
			\fi
			\global\firstcharfalse%
		\fi
}
\newcommand\ProcessDigit
{%
	\ifnum\lst@mode=\lst@Pmode%
		\iffirstchar%
				\global\startedbyadigittrue%
			\fi
			\global\firstcharfalse%
  \fi
}
\lst@AddToHook{Output}%
{%
	\ifstartedbyadigit%
		\def\lst@thestyle{\color{number}}%
	\fi
	\global\firstchartrue%
	\global\startedbyadigitfalse%
}
\newtoks\python@toks
\python@toks={language=Python,tabsize=4,frame=none,commentstyle=\itshape\color{comment},basicstyle=\footnotesize\ttfamily\color{text},keywordstyle=[1]\color{keyword},keywordstyle=[2]\color{builtin},keywordstyle=[3]\itshape\color{instance},stringstyle=\color{string},identifierstyle=\color{text},morestring=[d]{"""},showspaces=false,sensitive=true,showstringspaces=false,numbers=none,numbersep=5pt,numberstyle=\tiny\color{linenumber},breaklines=true,gobble=4,upquote=true,framexleftmargin=0mm,xleftmargin=12pt,escapechar=|,rulecolor=\color{black},keepspaces=true,procnamekeys={def,class},procnamestyle=\bfseries\color{definition},morekeywords=[1]{as},morekeywords=[2]{True,False},morekeywords=[3]{self,@property},alsoletter={0123456789.},alsodigit={.},SelectCharTable=%
}
\def\add@savedef#1#2{%
  \begingroup\lccode`?=#1\relax
  \lowercase{\endgroup
  \edef\@temp{%
    \noexpand\lst@DefSaveDef{\number#1}%
    \expandafter\noexpand\csname lsts@?\endcsname{%
      \expandafter\noexpand\csname lsts@?\endcsname\noexpand#2}%
  }}%
  \python@toks=\expandafter{\the\expandafter\python@toks\@temp}%
}
\count@=`0
\loop
  \add@savedef\count@\ProcessDigit
  \ifnum\count@<`9
  \advance\count@\@ne
\repeat
\count@=`A
\loop
  \add@savedef\count@\ProcessLetter
  \ifnum\count@<`Z
  \advance\count@\@ne
\repeat
\count@=`a
\loop
  \add@savedef\count@\ProcessLetter
  \ifnum\count@<`z
  \advance\count@\@ne
\repeat
\begingroup\edef\x{\endgroup
  \noexpand\lstdefinestyle{defaultpython}{\the\python@toks}
}\x
\makeatother

\definecolor{azure}{rgb}{0.0, 0.5, 1.0}
\definecolor{indiagreen}{rgb}{0.07, 0.53, 0.03}

\renewcommand{\textsw}[1]{\textsc{#1}}								
\newcommand{\textcl}[1]{\textsc{#1}}								

\newcommand{\python}{\textsw{Python}}

\newcommand{\prism}{\textsw{Prism}}

\newcommand{\modellink}{\textcl{ModelLink}}

\newcommand{\meraxes}{\textsw{Meraxes}}

\defcitealias{Meraxes}{M16}
\defcitealias{Qiu2019}{Q19}
\defcitealias{PRISM_ApJS}{V19}

\newcommand{\BLA}{Bayes linear approach}

\newcommand{\cov}{\mathrm{Cov}}							
\newcommand{\var}{\mathrm{Var}}							
\newcommand{\E}{\mathrm{E}}								

\newcommand{\mdvar}{$\var(\epsilon_{\mathrm{md,i}})$}   

\renewcommand{\vec}[1]{\bm{\mathrm{#1}}}

\makeatletter
\preto{\@verbatim}{\topsep=0.5\baselineskip  \partopsep=0.5\baselineskip}
\makeatother

\hypersetup{
	pdfauthor={Van der Velden et al.},%
    pdftitle={Ultra-fast model emulation with PRISM},
}

\received{22/11/2020}
\revised{04/01/2021}
\accepted{18/01/2021}
\submitjournal{ApJS}

\shorttitle{Ultra-fast model emulation with PRISM}
\shortauthors{Van der Velden et al.}

\begin{document}

\title{Ultra-fast model emulation with PRISM; analyzing the Meraxes galaxy formation model}

\correspondingauthor{Ellert van der Velden}
\email{evandervelden@swin.edu.au}

\author[0000-0002-1559-9832]{Ellert van der Velden}
\affiliation{Centre for Astrophysics and Supercomputing, Swinburne University of Technology, PO Box 218, Hawthorn, VIC 3122, Australia}
\affiliation{ARC Centre of Excellence for All Sky Astrophysics in 3 Dimensions (ASTRO 3D)}

\author[0000-0002-9636-1809]{Alan R. Duffy}
\affiliation{Centre for Astrophysics and Supercomputing, Swinburne University of Technology, PO Box 218, Hawthorn, VIC 3122, Australia}
\affiliation{ARC Centre of Excellence for All Sky Astrophysics in 3 Dimensions (ASTRO 3D)}

\author[0000-0002-5009-512X]{Darren Croton}
\affiliation{Centre for Astrophysics and Supercomputing, Swinburne University of Technology, PO Box 218, Hawthorn, VIC 3122, Australia}
\affiliation{ARC Centre of Excellence for All Sky Astrophysics in 3 Dimensions (ASTRO 3D)}

\author[0000-0002-3166-4614]{Simon J. Mutch}
\affiliation{ARC Centre of Excellence for All Sky Astrophysics in 3 Dimensions (ASTRO 3D)}
\affiliation{School of Physics, University of Melbourne, Parkville, VIC 3010, Australia}




\begin{abstract}
    We demonstrate the potential of an emulator-based approach to analyzing galaxy formation models in the domain where constraining data is limited.
    We have applied the open-source \python\ package \prism\ to the galaxy formation model \meraxes.
    \meraxes\ is a semi-analytic model, purposefully built to study the growth of galaxies during the Epoch of Reionization (EoR).
    Constraining such models is however complicated by the scarcity of observational data in the EoR.
    \prism's ability to rapidly construct accurate approximations of complex scientific models using minimal data is therefore key to performing this analysis well.
    
    This paper provides an overview of our analysis of \meraxes\ using measurements of galaxy stellar mass densities; luminosity functions; and color-magnitude relations.
    We demonstrate the power of using \prism\ instead of a full Bayesian analysis when dealing with highly correlated model parameters \textit{and} a scarce set of observational data.
    Our results show that the various observational data sets constrain \meraxes\ differently and do not necessarily agree with each other, signifying the importance of using multiple observational data types when constraining such models.
    Furthermore, we show that \prism\ can detect when model parameters are too correlated or cannot be constrained effectively.
    We conclude that a mixture of different observational data types, even when they are scarce or inaccurate, is a priority for understanding galaxy formation and that emulation frameworks like \prism\ can guide the selection of such data.
\end{abstract}

\keywords{methods: data analysis -- methods: numerical}



\section{Introduction}
\label{sec:Introduction}
Recent years have seen the establishment of a concordance cosmological model, called \textit{$\Lambda$CDM}, based on decades of increasingly sophisticated observations, with the universe being made up of ${\sim}75\%$ \textit{dark energy} and ${\sim}25\%$ mass (\textit{dark matter}, DM; and \textit{baryonic matter}) \citep{Planck15_13}.
Within the $\Lambda$CDM paradigm, a key transition occurs in the early universe, when sufficient ionizing photons have been produced to reionize the majority of the volume, called the \textit{Epoch of Reionization} (EoR; \citealt{EoR}).
Studying the EoR is quite challenging, as a limited number of observations is available due to the very high redshifts at which this event occurs, plus the opaque wall of neutral hydrogen along the line-of-sight.
Also, the formation of large-scale structures during the EoR is influenced by small-scale physics, such as star formation, making modelling the EoR rather complex.
In order to attempt to explain the EoR with galaxy formation, three different main types of approaches are used: N-body simulations; hydrodynamic simulations; and semi-analytic models.

Following the descriptions given in \citet{Lacey01}; \citet{Benson10} and \citet{Kuhlen12}, an \textit{N-body simulation} models the physics of large-scale galaxy structure, by directly evolving a large number of particles solely based on the fundamental equations of gravitation.
Such simulations are very successful (like the \textit{Millennium simulation}; \citealt{Millennium}), as DM particles are known to only interact gravitationally and make up the majority of the total mass, leaving baryonic matter with only a small contribution.

\textit{Hydrodynamic simulations} (e.g., \citealt{Schaye10,Illustris,EAGLE}) involve, as the name already suggests, explicit calculations of the hydrodynamic forces together with gravity.
The inclusion of hydrodynamics means that baryons play a more explicit role in structure formation, especially at small scales, because baryons can radiatively lose energy and thereby collapse to higher densities.
Hence, to follow their evolution, the required resolution timescales are much smaller as well as being more complex to simulate in general.
Therefore, hydrodynamic simulations tend to be much more computationally expensive than N-body simulations.

Finally, \textit{semi-analytic models} (SAMs; e.g., \citealt{Galform,Croton06,Henriques13,Somerville15,SAGE,Meraxes,DRAGONS-X,Shark}) use approximations of the involved physics to model galaxy formation.
SAMs are commonly applied to a static halo merger tree that is obtained from an N-body simulation, and the galaxy formation physics is treated in a phenomenological way.
In practice, this means that there are typically far more parameters in the model that can be tuned than in hydrodynamic simulations, but SAMs are also far less computationally expensive.
This allows SAMs to explore different evolution tracks much quicker, providing broader insight into the underlying physics.
The disadvantage, however, is that a higher degree of approximation is used, which can give rise to unforeseen artifacts and unknown variances.
In order to find out to what extent this has an effect on the accuracy and reliability of a SAM, the model needs to be thoroughly analyzed. 
While this paper explores a means to efficiently and thoroughly explore SAMs, we note that the technique could be similarly valuable in hydrodynamic simulations and their parametrizations.

Commonly, a combination of \textit{Markov chain Monte Carlo} (MCMC) methods and \textit{Bayesian statistics} (e.g., \citealt{BayesianBook,gelman2014bayesian}) is employed when performing model parameter estimations, as performed by \citet{Henriques13,Martindale17,Henriques19} for galaxy formation SAMs.
This is used to obtain an accurate approximation of the posterior \textit{probability distribution function} (PDF), which allows for regions in parameter space that compare well to the available data to be identified.
As a consequence, MCMC and Bayesian statistics tend to be used for analyzing (exploring the behavior of) a model as well, even though obtaining the posterior PDF is an inherently slow process.
However, \citet{PRISM_ApJS}, hereafter \citetalias{PRISM_ApJS}, argued that an accurate approximation of the posterior PDF is not required for analyzing a model, but the converging process towards the PDF is required instead.
\citetalias{PRISM_ApJS} therefore proposed the combination of the \textit{\BLA} \citep{Goldstein99,Goldstein00,BLA} and the \textit{emulation technique} \citep{Sacks89,Currin91,Oakley02,O'Hagan06} with \textit{history matching} \citep{Raftery95,Craig96,Craig97,Kennedy01,Goldsteinetal06} as an alternative.
A combination of these techniques has been applied to galaxy formation SAMs before \citep{Bower10,Vernon10,Vernon14,Rodriques17}, showing their value.

In this work, we use an open-source \python\ implementation of these techniques, called \prism\ \citep{PRISM_ApJS,PRISM_JOSS}, to explore one such semi-analytic model; \meraxes\ \citep{Meraxes}.
Unlike other SAMs, like \textsw{SAGE} \citep{SAGE} and \textsw{Shark} \citep{Shark}, \meraxes\ walks through the aforementioned static halo trees of an N-body simulation \textit{horizontally} instead of \textit{vertically}.
This means that for each \textit{snapshot} (time instance), all halos at that snapshot within all trees are processed simultaneously, such that interactions between the different trees can be taken into account.
For \meraxes, this is necessary, as it was purposefully built to self-consistently couple the galaxy formation to the reionization history at high redshifts ($z\geq 5$).
This makes the model suitable to obtain more insight into reionization, but it also makes it much slower to explore due to the coupling mechanism requiring many additional calculations for each snapshot.
It is therefore crucial that we use a framework like \prism\ to efficiently explore and analyze \meraxes.

This work is structured in the following way: In \ref{sec:Emulation}, we give a short overview of how the \BLA, emulation technique and history matching can be used to analyze models.
Then, using the \prism\ framework, we analyze the \meraxes\ model and discuss our findings in \ref{sec:M16}.
After that, we explore a modified version of \meraxes\ by \citet{Qiu2019} in \ref{sec:Q19}, and compare our results with theirs.
Finally, in \ref{sec:Conclusions}, we give a summary of the results in this work and discuss the potential implications they may have.

\section{Model analysis \& emulation}
\label{sec:Emulation}
In this section, we give a short overview of how emulation in \prism\ works and how it can be used to analyze scientific models efficiently.
Note that this section is meant for those seeking a basic understanding of the methodology and terminology used throughout this work.
We refer to \citet{BLA} and \citet{Vernon10} for further details on the \BLA, emulation technique and history matching; or to \citetalias{PRISM_ApJS} for their specific use in \prism.

The basics of the \textit{emulation technique} is as follows.
Suppose that we have a model that takes a vector of input parameters $\vec{x}$, and whose output is given by the hypothetical function $f(\vec{x})$.
As $f(\vec{x})$ is both continuous and defined over a closed input interval, we can apply the \textit{Stone-Weierstrass theorem} \citep{Stone48}, which states that such a function can be uniformly approximated by a polynomial function as closely as desired.
Using this, we can say that the output $i$ of $f(\vec{x})$, $f_i(\vec{x})$, is given by
\begin{align}
\label{eq:fx}
    f_i(\vec{x}) &= \sum_j\beta_{ij}g_{ij}(\vec{x}_{A,i})+u_i(\vec{x}_{A,i})+w_i(\vec{x}),
\end{align}
with $\beta_{ij}$ unknown scalars; $g_{ij}$ deterministic functions of $\vec{x}$; $u_i(\vec{x})$ a weakly stochastic process with constant variance; and $w_i(\vec{x})$ all remaining variance.
The vector $\vec{x}_{A,i}$ represents the vector of \textit{active} input arguments for output $i$; those values that are considered to have a non-negligible impact on the output.

The goal of an emulator is to find out what the form of this function $f(\vec{x})$ is for a given model.
However, as model evaluations tend to be complex and very time-consuming, we would like to avoid evaluating the model millions of times.
To solve this problem, we use the \textit{\BLA}, which can be seen as an approximation of a full Bayesian analysis that uses expectations instead of probabilities as its main output.
By using the \BLA, for a given vector of known model realizations $D_i=\left(f_i(\vec{x}^{(1)}), f_i(\vec{x}^{(2)}), \ldots, f_i(\vec{x}^{(n)})\right)$, we can calculate what the \textit{adjusted expectation} and \textit{adjusted variance} values are for a given output $i$:
\begin{align}
\label{eq:adj_exp}
    \begin{split}
        &\E_{D_i}(f_i(\vec{x})) = \\
        &\quad\E(f_i(\vec{x}))+\cov\left(f_i(\vec{x}), D_i\right)\cdot\var(D_i)^{-1}\cdot\left(D_i-\E(D_i)\right),
    \end{split}\\
\label{eq:adj_var}
    \begin{split}
        &\var_{D_i}(f_i(\vec{x})) = \\
        &\quad\var(f_i(\vec{x}))-\cov(f_i(\vec{x}), D_i)\cdot \var(D_i)^{-1}\cdot \cov(D_i, f_i(\vec{x})),
    \end{split}
\end{align}
with $\E(f_{i}(\vec{x}))=\sum_j\E(\beta_{ij})g_{ij}(\vec{x}_{A,i})$ the prior expectation of $f_i(\vec{x})$; $\cov(f_i(\vec{x}), D_i)$ the vector of covariances between the unknown output $f_i(\vec{x})$ and all known outputs $D_i$; and $\var(D_i)$ the $n\times n$ matrix of covariances between all known outputs with elements $\var_{jk}(D_i)=\cov(f_i(\vec{x}^{(j)}), f_i(\vec{x}^{(k)}))$.

\ref{eq:adj_exp} and \ref{eq:adj_var} can be used to update our beliefs about the function $f_i(\vec{x})$ given a vector of known outputs $D_i$.
With our updated beliefs, we can determine the collection $\mathcal{X}^*$, which contains those input parameters $\vec{x}$ that give an `acceptable' match to the observations $z$\footnote{Not to be confused with redshift} when evaluated in $f(\vec{x})$, including the unknown `best' vector of input parameters $\vec{x}^*$.
The process of obtaining this collection $\mathcal{X}^*$ is called \textit{history matching}, which can be seen as the equivalent to \textit{model calibration} when performing a full Bayesian analysis.
History matching is achieved through evaluating functions called \textit{implausibility measures} \citep{Craig96,Craig97}, which have the following form:
\begin{align}
\label{eq:impl_sq}
    I_i^2(\vec{x}) &= \frac{\left(\E_{D_i}(f_i(\vec{x}))-z_i\right)^2}{\var_{D_i}(f_i(\vec{x}))+\var(\epsilon_{\mathrm{md}, i})+\var(\epsilon_{\mathrm{obs}, i})},
\end{align}
with $\E_{D_i}(f_i(\vec{x}))$ the adjusted emulator expectation (\ref{eq:adj_exp}); $\var_{D_i}(f_i(\vec{x}))$ the adjusted emulator variance (\ref{eq:adj_var}); \mdvar\ the model discrepancy variance; and $\var(\epsilon_{\mathrm{obs}, i})$ the observational variance.
Here, the \textit{model discrepancy variance} \citep{Kennedy01,Vernon10} is a measure of the (un)certainty that the output of the model is correct, given its intrinsic implementation and functionality (e.g., approximations; stochasticity; missing features; etc.)
As this variance can be rather challenging to properly calculate, it is usually estimated using a justified assumption.

For a given input parameter set $\vec{x}$, the corresponding implausibility value $I_i(\vec{x})$ tells us how (un)likely it is that we would view the match between the model output $f_i(\vec{x})$ and the observational data $z_i$ as acceptable or \textit{plausible}, in terms of the standard deviation $\sigma$.
The higher the implausibility value, the more unlikely it is that we would consider $\vec{x}$ to be part of the collection $\mathcal{X}^*$.
Because the implausibility measure is both unimodal and continuous, we can use the $3\sigma$-rule given by \citet{Pukelsheim94} to show that $95\%$ of its probability must lie within $\pm3\sigma$ ($I_i(\vec{x})\le3$).
Values higher than $3$ would usually mean that the proposed input parameter set $\vec{x}$ should be discarded, but we show later in this work that this is not always necessary in order to heavily reduce parameter space.

\begin{figure*}
\begin{center}
	\subfloat[Initial Gaussian emulator with $5$ model evaluations.]{\label{subfig:gaussian_0D_1}\includegraphics[width=\textwidth]{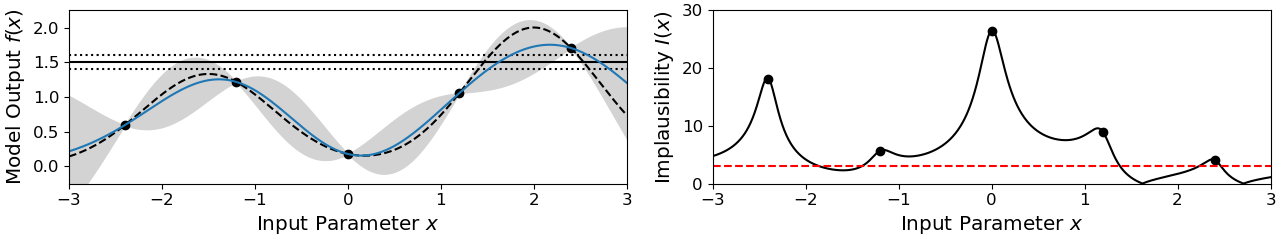}}\\
    \subfloat[Updated Gaussian emulator with $11$ model evaluations.]{\label{subfig:gaussian_0D_2}\includegraphics[width=\textwidth]{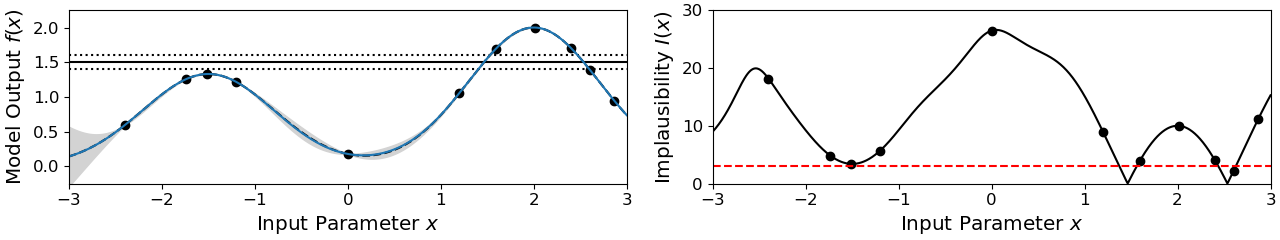}}
	\caption{Emulator of two simple Gaussians, defined as $f(x)=2\cdot\exp\left(-(2-x)^2\right)+1.33\cdot\exp\left(-(-1.5-x)^2\right)$.
	\textbf{Left:} Gaussian model $f(x)$ (\textbf{dashed}), model evaluations $D$ (\textbf{dots}), emulator $\E_D(f(x))$ (\textbf{solid}), emulator uncertainty $\E_D(f(x))\pm 3\sqrt{\var_D(f(x))}$ (\textbf{shaded}), observational data with $2\sigma$ errors (\textbf{horizontal lines}).
	\textbf{Right:} Implausibility values $I(x)$ (\textbf{solid}) with cut-off $I_{\mathrm{cut,1}}=3$ (\textbf{dashed}).
    \textit{Reproduced from \citetalias{PRISM_ApJS}.}}
    \label{fig:gaussian_0D}
\end{center}
\end{figure*}

To show how the theory above can be applied to a model, we have created an emulator of a simple Gaussian function, given in \ref{subfig:gaussian_0D_1}.
On the left in the figure, we show the model output function $f(x)$ as the dashed line, which is usually not known but shown here for convenience.
This model has been evaluated a total of five times, represented by the black dots.
Using these evaluations, we have created an emulator using \ref{eq:adj_exp} and \ref{eq:adj_var}, given by the solid line and the shaded area.
This shaded area shows the $3\sigma$-confidence interval, with $\sigma=\sqrt{\var_D(f(x))}$.
Finally, we added a single comparison data point, given as the horizontal line with its $2\sigma$ confidence interval.

Using all of this information, we can calculate what the implausibility values are for all values of $x$ using \ref{eq:impl_sq}, which is shown on the right in \ref{subfig:gaussian_0D_1}.
If we state that we consider values of $I(x)\le3$ to be acceptable, as indicated by the dashed line, then we can see that there are only a few regions in parameter space that satisfy this condition and thus require further analysis.
After evaluating the model six more times in these regions and updating our beliefs, we obtain the plots in \ref{subfig:gaussian_0D_2}.
We can now see that the emulator solely focused on the important parts of parameter space, where it has been greatly improved, and that the implausibility values are only acceptable for $x=1.5$ and $x=2.5$, which is as expected.

When making emulators of models that are more complex than the one shown in \autoref{fig:gaussian_0D}, it is very likely that the model has more than a single output (and thus multiple comparison data points).
In this scenario, every model output has its own defined implausibility measure, which are evaluated and combined together.
In order for an input parameter set $\vec{x}$ to be considered plausible, all implausibility measures must meet a specific criterion.
This criterion, called the \textit{implausibility cut-off}, is given by
\begin{align}
\label{eq:impl_cut}
    I_{\mathrm{max}, n}(\vec{x}) &\le I_{\mathrm{cut}, n},
\end{align}
with $I_{\mathrm{cut, n}}$ being the maximum value that the $n$th highest implausibility value $I_{\mathrm{max, n}}(\vec{x})$ is allowed to have, for a given $\vec{x}$.
In cases where there are many model outputs, it can sometimes be desirable to ignore the first few highest implausibility measures regardless of their values.
In this scenario, we apply so-called \textit{implausibility wildcards} to the emulator, which allows us to analyze a model in a slower fashion by ignoring as many implausibility measures as there are wildcards.
As a higher implausibility value implies a more constraining data point, this effectively removes the heavier constraints from the analysis.

By performing history matching iteratively, a process called \textit{refocusing}, we can remove parts of parameter space based on the implausibility values of evaluated input parameter sets.
This in turn leads to a smaller parameter space to evaluate the model in, and thus a higher density of parameter sets to update our beliefs with.
As this leads to a more accurate emulator, we can progressively make a better approximation of the model with each iteration.

The concepts discussed in this section allow the \prism\ framework to quickly make accurate approximations of models in those parts of parameter space that are important, and provide us with insights into the behavior of the model.
In the remainder of this work, we use its unique features to analyze the \meraxes\ model in various different ways, as shown in \ref{sec:M16} and \ref{sec:Q19} using \prism\ version 1.3.0.
For more detailed information on \prism, see \citetalias{PRISM_ApJS}.

\section{Analyzing the Meraxes model}
\label{sec:M16}
In this section, we discuss the results of our analysis of the \meraxes\ galaxy formation model by \citet{Meraxes}, hereafter \citetalias{Meraxes}.
These results are based on two emulators made with \prism.
The emulators only differ in the constraining observational data that was used, and are described below.

The emulator uses a \modellink\ subclass (a \python\ class that wraps a model such that \prism\ can interface with it) that maps the \textit{stellar mass function} (SMF) number density values from its logarithmic scale to an $\arctan$ function, while also mapping the associated SMF data errors accordingly as well.
This results in all SMF data values to be in the range $[\nicefrac{-\pi}{2}, \nicefrac{\pi}{2}] = [-1.571, 1.571]$, which was used for all subsequent emulators in this work as well.
The reason we use an $\arctan$ function here is to limit the range of orders of magnitude for the data values, decreasing the possibility of artifacts due to floating point errors in the emulation process.
The \modellink\ subclass uses the SMF parameter values as described in Table 1 in \citetalias{Meraxes} with appropriate ranges, which are given in \autoref{tab:M16_par}.
We used the SMFs presented in \citet{Song2016} at redshifts $z=[7, 6, 5]$ to constrain the parameters for a combined total of $24$ data points.
Note that because of this, in the following, whenever we refer to the SMF, we specifically mean the SMF at these redshifts.
As the \meraxes\ model is expected to be reasonably well constrained in this redshift range, we assumed a static model discrepancy variance \mdvar\ of $10^{-4}$ for all data points in this section.

\begin{table}
    \centering
    \begin{tabular}{|c|c|c|c|}
    \hline
        \textbf{Name} & \textbf{Min} & \textbf{Max} & \textbf{Estimate}\\
    \hline
        \texttt{MergerBurstFactor} & $0.0$ & $1.0$ & $0.57$ \\
        \texttt{SfCriticalSDNorm} & $0.0$ & $1.0$ & $0.2$ \\
        \texttt{SfEfficiency} & $0.0$ & $1.0$ & $0.03$ \\
        \texttt{SnEjectionEff} & $0.0$ & $1.0$ & $0.5$ \\
        \texttt{SnEjectionNorm} & $0.0$ & $100.0$ & $70.0$ \\
        \texttt{SnEjectionScaling} & $0.0$ & $10.0$ & $2.0$ \\
        \texttt{SnReheatEff} & $0.0$ & $10.0$ & $6.0$ \\
        \texttt{SnReheatNorm} & $0.0$ & $100.0$ & $70.0$ \\
        \texttt{SnReheatScaling} & $0.0$ & $10.0$ & $0.0$ \\
    \hline
    \end{tabular}
    \caption{The parameters used by \prism\ for calculating the SMF in \meraxes, where all other (potentially important) parameters were set to their default values at all times.
    The boundaries (where possible) and estimates were obtained from Table 1 in \citetalias{Meraxes}, with missing boundary values being chosen within reason.
    The names in the \textbf{first column} indicate the name that these parameters have in \meraxes.}
    \label{tab:M16_par}
\end{table}

The nine parameters shown in \autoref{tab:M16_par} can be subdivided into two different groups; namely those related to the star formation (\texttt{MergerBurstFactor}; \texttt{SfCriticalSDNorm}; and \texttt{SfEfficiency}) and the supernova feedback (\texttt{SnEjection\dots}\ and \texttt{SnReheat\dots}).
The star formation parameters determine the star formation burst that happens after a galaxy/halo merger (\texttt{MergerBurstFactor}); the critical surface density required for star formation (\texttt{SfCriticalSDNorm}); and the efficiency with which available cold gas is converted into stars (\texttt{SfEfficiency}).
Therefore, all three of these parameters directly influence the stellar mass of a galaxy/halo.
On the other hand, the \texttt{SnEjection\dots}\ and \texttt{SnReheat\dots}\ parameters are each a group of three free parameters that determine a single physical quantity in \meraxes; the supernova energy coupling efficiency and the mass loading factor, respectively.
Since the supernova feedback heavily influences the amount of (cold) gas that is available for star formation, they are very important for the model predictions.

\subsection{Testing the waters with mock data}
\label{subsec:M16_mock}
Before exploring the \meraxes\ model, we first had to determine whether \prism\ was capable of handling such a complex non-linear model effectively.
In order to do this, we created mock comparison data for which we knew the model realization $\vec{x}^*$.
For convenience, we set the values of $\vec{x}^*$ to be the estimates given in \autoref{tab:M16_par}.
Using this parameter set $\vec{x}^*$, we evaluated \meraxes\ and retrieved the output values at the same $24$ points as are given by \citet{Song2016}, creating the mock data set.
As mock data does not have an observational error associated with it, we used the square root of the model discrepancy variance \mdvar\ as the error.
Finally, as it is impossible for a model to make `perfect' predictions, we also perturbed the mock data values by their own errors using a normal distribution.

\begin{table}
    \centering
    \begin{tabular}{|c|r|l|c|l|}
    \hline
        \multicolumn{1}{|c|}{$i_{\mathrm{emul}}$} &
        \multicolumn{1}{|c|}{$n_{\mathrm{eval}}$} & 
        \multicolumn{1}{|c|}{$I_{\mathrm{cut,n}}$} & 
        \multicolumn{1}{|c|}{$n_{\mathrm{wild}}$} &
        \multicolumn{1}{|c|}{$f_{\mathrm{space}}$} \\
    \hline
        $1$ & $500$ & $[4.0, 3.5, 3.2, 3.0]$ & $3$ & $1.32\%$ \\
        $2$ & $712$ & $[4.0, 3.5, 3.2, 3.0]$ & $2$ & $0.144\%$ \\
        $3$ & $1,036$ & $[4.0, 3.5, 3.2, 3.0]$ & $2$ & $0.0748\%$ \\
        $4$ & $808$ & $[4.0, 3.5, 3.2, 3.0]$ & $1$ & $0.0262\%$ \\
    \hline
    \end{tabular}
    \caption{Statistics for the \meraxes\ emulator with mock data.
    The \textbf{first column} specifies the emulator iteration $i_{\mathrm{emul}}$ this row is about.
    The \textbf{next three columns} provide the number of model evaluations $n_{\mathrm{eval}}$; the non-wildcard implausibility cut-offs $I_{\mathrm{cut,n}}$; and the number of implausibility wildcards $n_{\mathrm{wild}}$ used for this emulator iteration.
    Finally, the \textbf{last column} gives the fraction of parameter space remaining $f_{\mathrm{space}}$ after this emulator iteration was analyzed.}
    \label{tab:M16_mock_stats}
\end{table}

Using this mock data, we created an emulator of \meraxes, whose statistics are shown in \autoref{tab:M16_mock_stats}.
Since we are simply testing if \prism\ works properly, we purposefully chose high implausibility cut-offs and several implausibility wildcards for this emulator, such that it would converge slower.
As described earlier, every implausibility wildcard causes the next highest implausibility measure to be ignored when evaluating the emulator.
Despite this however, the emulator converged quickly, demonstrating that small variances have a much bigger impact than the conservative choices for the algorithm.
Because the observational variance and model discrepancy variance are the same, this may have a negative effect on the accuracy of the emulator (cf., \citetalias{PRISM_ApJS}), so we kept in mind that this might be an issue in the case that the emulator is incorrect.

\begin{figure}
    \centering
    \subfloat{\includegraphics[width=0.49\linewidth]{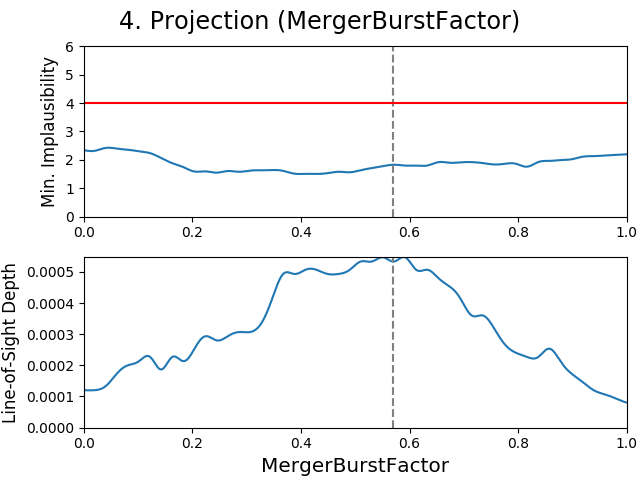}}
    \subfloat{\includegraphics[width=0.49\linewidth]{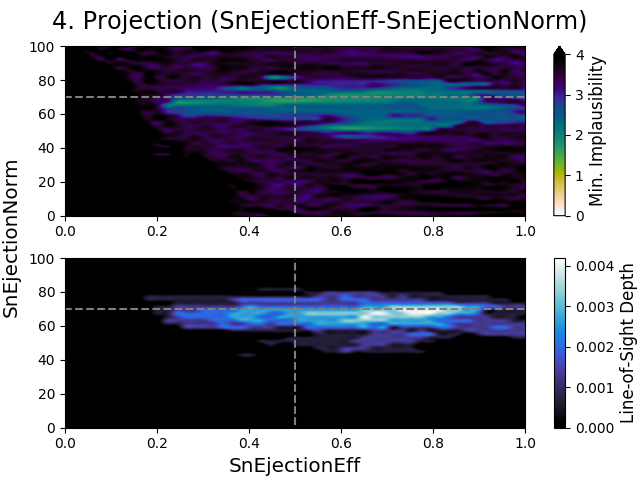}} \\
    \subfloat{\includegraphics[width=0.49\linewidth]{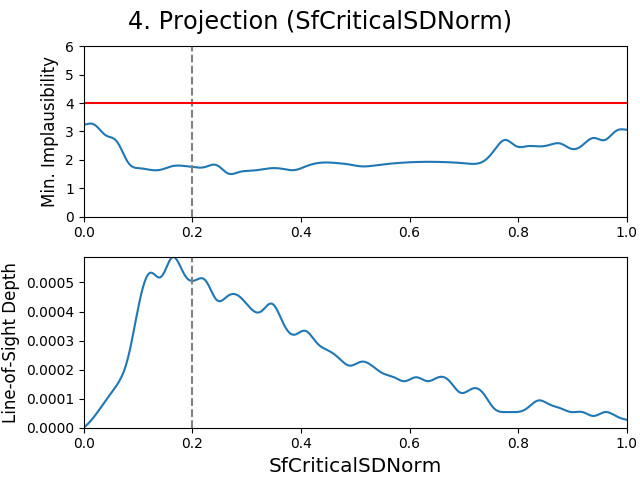}}
    \subfloat{\includegraphics[width=0.49\linewidth]{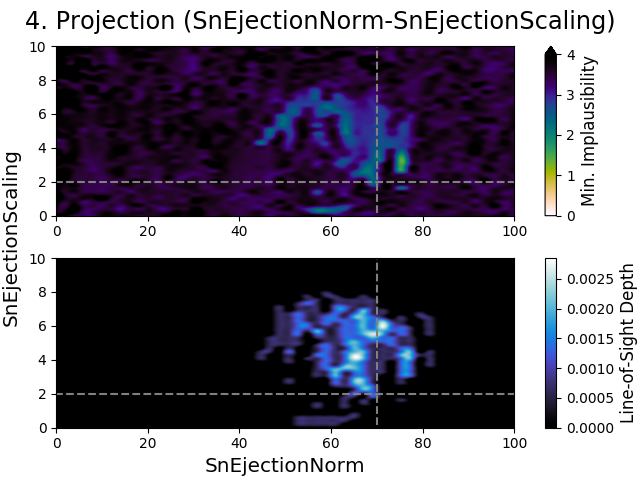}} \\
    \subfloat{\includegraphics[width=0.49\linewidth]{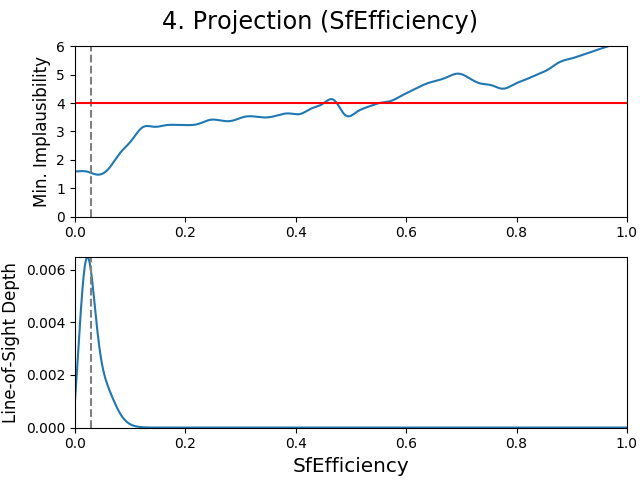}}
    \subfloat{\includegraphics[width=0.49\linewidth]{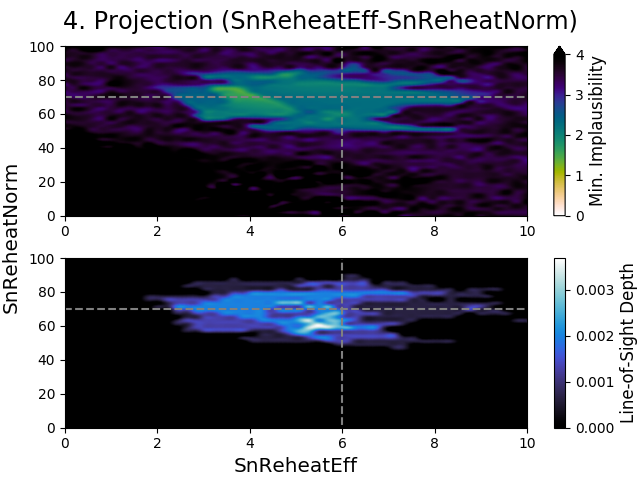}}
    \caption{Projection figures of the \meraxes\ emulator with mock data at $i_{\mathrm{emul}}=4$.
    The \textbf{dashed lines} show the estimated value of the corresponding parameter as given in \autoref{tab:M16_par}.
    \textbf{Left:} Three 2D projection figures showing the behavior of the parameters related to the star formation.
    \textbf{Right:} Three 3D projection figures showing the behavior of and correlations between the parameters related to the supernova feedback.}
    \label{fig:M16_mock_results}
\end{figure}

In order to study the behavior of an emulator, \prism\ creates a series of 2D and 3D \textit{projection figures}, which for this emulator are shown in \autoref{fig:M16_mock_results}.
As described by \citetalias{PRISM_ApJS}, a 3D projection figure consists of two subplots for every combination of two active model parameters.
Each subplot is created by analyzing a grid of $25\times25$ points for the plotted parameters, where a Latin-Hypercube design \citep{McKay79} of $1,500$ samples is used for the remaining parameter values in every grid point.
All results are then processed to yield a single result per grid point that is independent of the non-plotted parameters.
In every projection figure, the top subplot shows the minimum implausibility value (i.e., $\min(I_{\mathrm{max,n}})$ with $n$ the first non-wildcard cut-off) that can be reached, whereas the bottom subplot shows the fraction of samples (``line-of-sight depth'') that is plausible.
An easy way to distinguish these plots is that the former shows where the \textit{best} plausible samples can be found, while the latter shows where the \textit{most} plausible samples are.
A 2D projection is similar to a 3D projection, but only a single model parameter is plotted, and is more comparable to a \textit{marginalized likelihood/density plot} often used in Bayesian statistics.

A combination of 2D and 3D projections can be used to study many properties of the \meraxes\ model.
On the left in \autoref{fig:M16_mock_results}, we show the 2D projections of the three star formation parameters.
These three projections show us quite clearly where the best parameter values can be found, especially for the star formation efficiency \texttt{SfEfficiency}, which all agree very well with their parameter estimates (the dashed lines).
Although we know that the \meraxes\ model is capable of generating a model realization that corresponds to the data, as mock data was used instead of observational data, this result is still somewhat surprising.
After all, almost the entire parameter range is still considered to be plausible for the \texttt{MergerBurstFactor} and \texttt{SfCriticalSDNorm} parameters, while the favored parameter values are very obvious.
This shows that while \prism\ is quite conservative with removing plausible space despite low variances, it can still reach a result quickly, only requiring $3,056$ model evaluations to reduce plausible parameter space by almost a factor of $4,000$.

On the right of \autoref{fig:M16_mock_results}, we show the 3D projections of the free parameters related to the supernova feedback.
As each of these supernova feedback parameters are comprised of three free parameters, we expect the free parameters to be heavily correlated with each other.
As demonstrated by these figures, this is indeed the case.
While \prism\ can come to a good estimate of where the best parameter values are, it cannot pinpoint them nearly as easily as the star formation parameters.
This would either require more model evaluations or more data points, where the latter would hopefully provide more constraining information for these free parameters.

\subsection{Investigating the Meraxes model calibration}
\label{subsec:M16_SMF}
Now that we have established that \prism\ is indeed capable of retrieving the proper parameters with reasonable accuracy and a low number of model evaluations, we now use the observational data from \citet{Song2016}.
Apart from the different comparison data values, we kept all other variables, like the implausibility cut-offs and model discrepancy variance, as similar as possible.
This allows us to compare the results of this emulator with the mock data one we studied previously.
The statistics of this emulator are shown in \autoref{tab:M16_SMF_stats}.

\begin{table}
    \centering
    \begin{tabular}{|c|r|l|c|l|}
    \hline
        \multicolumn{1}{|c|}{$i_{\mathrm{emul}}$} &
        \multicolumn{1}{|c|}{$n_{\mathrm{eval}}$} & 
        \multicolumn{1}{|c|}{$I_{\mathrm{cut,n}}$} & 
        \multicolumn{1}{|c|}{$n_{\mathrm{wild}}$} &
        \multicolumn{1}{|c|}{$f_{\mathrm{space}}$} \\
    \hline
        $1$ & $500$ & $[4.0, 3.5, 3.2, 3.0]$ & $3$ & $2.29\%$ \\
        $2$ & $1,189$ & $[4.0, 3.5, 3.2, 3.0]$ & $2$ & $0.149\%$ \\
        $3$ & $878$ & $[4.0, 3.5, 3.2, 3.0]$ & $1$ & $0.00270\%$ \\
    \hline
    \end{tabular}
    \caption{Statistics for the \meraxes\ emulator using SMF data.
    The \textbf{first column} specifies the emulator iteration $i_{\mathrm{emul}}$ this row is about.
    The \textbf{next three columns} provide the number of model evaluations $n_{\mathrm{eval}}$; the non-wildcard implausibility cut-offs $I_{\mathrm{cut,n}}$; and the number of implausibility wildcards $n_{\mathrm{wild}}$ used for this emulator iteration.
    Finally, the \textbf{last column} gives the fraction of parameter space remaining $f_{\mathrm{space}}$ after this emulator iteration was analyzed.}
    \label{tab:M16_SMF_stats}
\end{table}

When comparing the emulator statistics given in \autoref{tab:M16_SMF_stats} with those in \autoref{tab:M16_mock_stats}, we can see that both emulators behaved very similarly in iterations $1$ and $2$.
However, because we noted that the mock data emulator barely changed when we did not decrease the number of implausibility wildcards, we decided to use the same implausibility parameters for iteration $3$ as we did for iteration $4$ in \autoref{tab:M16_mock_stats}.
However, unlike the mock data emulator, this emulator vastly reduced its plausible parameter space, doing so by more than a factor $50$.
This can be an indication that either there are a select few data points that are heavily constraining the model (as they would be ignored when using more implausibility wildcards), or that the model cannot produce any realizations that can explain the data reasonably well.
Given that the former trend was not observed in the mock data emulator, even though it would apply there as well, we expect that the latter is playing a role here.

\begin{figure}
    \centering
    \subfloat{\includegraphics[width=0.49\linewidth]{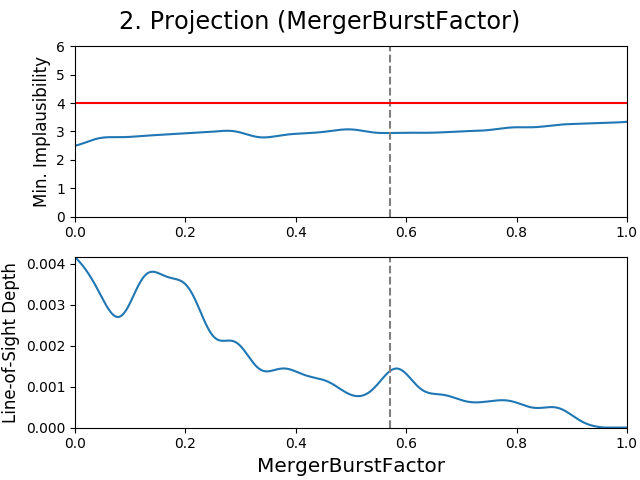}}
    \subfloat{\includegraphics[width=0.49\linewidth]{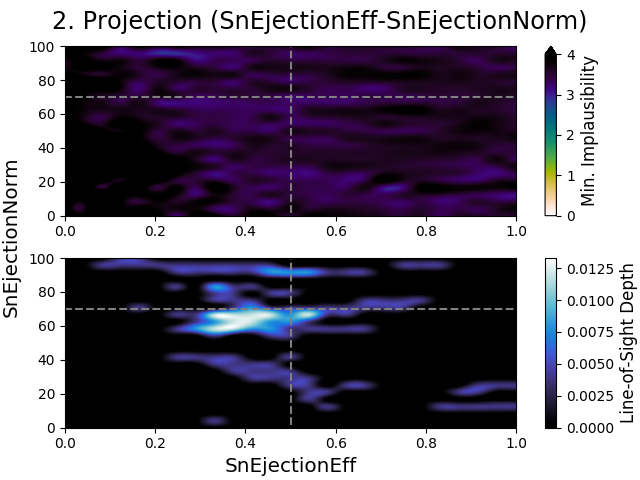}} \\
    \subfloat{\includegraphics[width=0.49\linewidth]{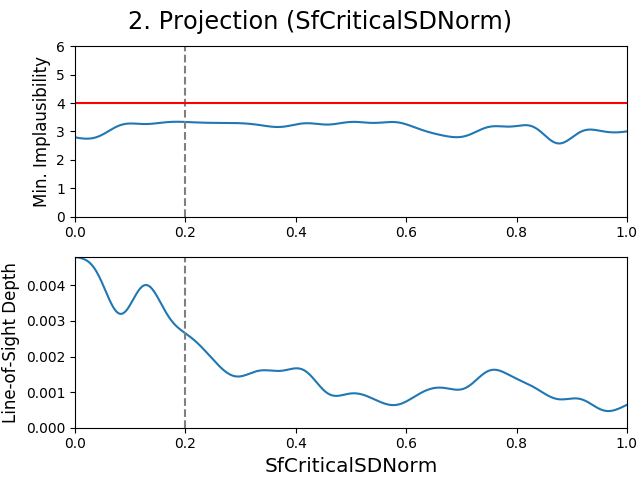}}
    \subfloat{\includegraphics[width=0.49\linewidth]{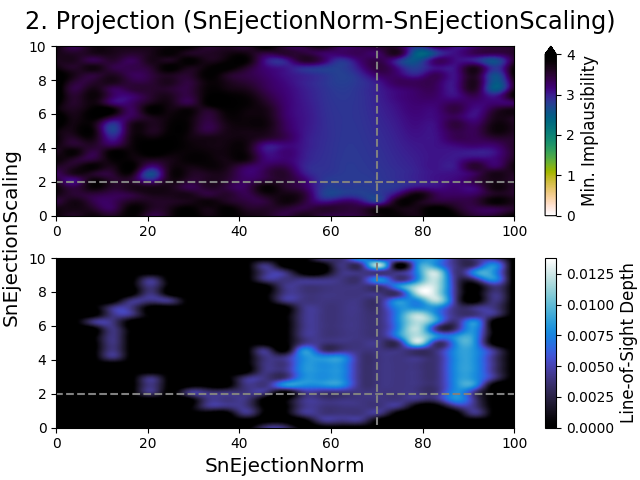}} \\
    \subfloat{\includegraphics[width=0.49\linewidth]{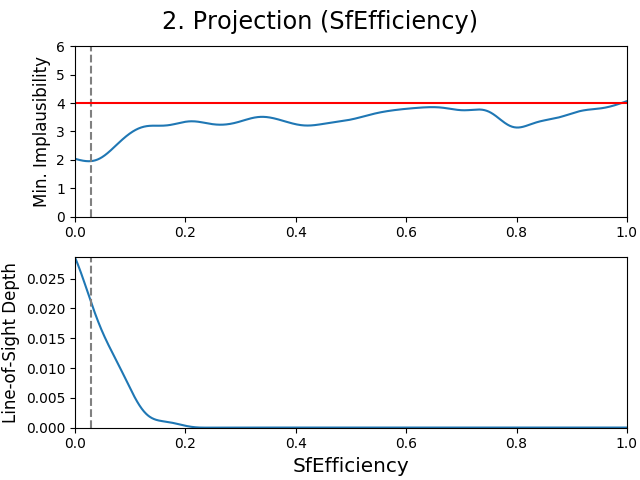}}
    \subfloat{\includegraphics[width=0.49\linewidth]{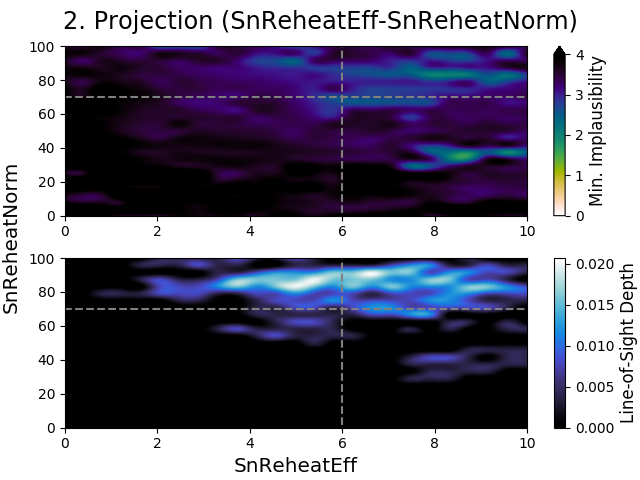}}
    \caption{Projection figures of the \meraxes\ emulator using SMF data at iteration $i_{\mathrm{emul}} = 2$.
    The \textbf{dashed lines} show the estimated value of the corresponding parameter as given in \autoref{tab:M16_par}.
    \textbf{Left:} Three 2D projection figures showing the behavior of the parameters related to the star formation.
    \textbf{Right:} Three 3D projection figures showing the behavior of and correlations between the parameters related to the supernova feedback.}
    \label{fig:M16_SMF_results}
\end{figure}

In \autoref{fig:M16_SMF_results}, we show the projection figures of this emulator for iteration $2$.
We show the iteration $2$ projections here instead of the iteration $3$ ones, as the latter show much less structure than the former.
This is another indication that the \meraxes\ model might be incapable of producing realizations that fit the data, as a lack of structure is usually caused by no areas of parameter space being favored over others, thus creating a tiny hypercube shell of plausible samples.

From the iteration $2$ projections, we can see that the star formation parameters, specifically the \texttt{MergerBurstFactor} and \texttt{SfCriticalSDNorm}, are not agreeing with their own estimates.
Instead, it appears that they `prefer' much lower values than their estimates indicate.
The star formation efficiency \texttt{SfEfficiency} does agree with its estimate, but not nearly with the same confidence as it did in \autoref{fig:M16_mock_results}.
We kept in mind here that this specific emulator iteration has a plausible space that is five times larger than the aforementioned emulator's plausible space, but the plausible samples are still much more spread out.
This is especially evident when comparing the 2D projection for \texttt{SfEfficiency} in both figures with each other.

However, when we examine the 3D projections in \autoref{fig:M16_SMF_results}, we can clearly see that the range of plausible values is very large.
The minimum implausibility plots show barely any structure (as this gets even worse for iteration $3$, we do not show these figures here), unlike the corresponding projections in \autoref{fig:M16_mock_results}.
Furthermore, the parameters are not consistent with their estimates at all.
Because we noted earlier that the supernova feedback parameters used here are in fact groups of three free parameters that combined determine a single physical process, this might be an indication that the estimates are significantly perturbed due to correlation errors.

Using the results that we have collected thus far, mainly that the parameters do not converge, we conclude that the manually calibrated parameter values, as given by \citetalias{Meraxes} and shown in \autoref{tab:M16_par}, are likely to be biased and thus non-optimal.
This conclusion is indeed implied by \citetalias{Meraxes}, whose caption of the parameter estimates in their Table 1 states that the \textit{``values were constrained to visually reproduce the observed evolution in the galaxy stellar mass function between $z=5$ and $z=7$''}.
Furthermore, even though we have shown that \prism\ is very capable of constraining \meraxes\ quickly, the free parameters for the energy coupling efficiency \texttt{SnEjection\dots}\ and the mass loading factor \texttt{SnReheat\dots}\ are very hard to constrain properly using the SMF at $z = [7, 6, 5]$, as shown by the 3D projections in \autoref{fig:M16_mock_results}.
\citet{Qiu2019} came to a similar conclusion, and decided to make some modifications to the \meraxes\ model and use luminosity data in order to attempt to constrain the star formation and supernova feedback parameters.
In the next section, we use their modified \meraxes\ model with \prism\ to see whether this resolves the problem, and discover what else we can learn from its analysis.

\section{Studying a modified Meraxes}
\label{sec:Q19}
In this section, we discuss the results of the emulators that were made using the methodology and parameter values for \meraxes\ as reported in \citet{Qiu2019}, hereafter \citetalias{Qiu2019}.
In the following, to avoid confusion, we refer to the two different versions of \meraxes\ using the corresponding work in which they are described (e.g., `\citetalias{Meraxes} \meraxes' and `\citetalias{Qiu2019} \meraxes'), or simply refer to it as `\meraxes' when we describe the model in general.

The \citetalias{Qiu2019} \meraxes\ model is substantially modified from the \citetalias{Meraxes} \meraxes\ model we explored in \ref{sec:M16}.
The most significant difference between the two \meraxes\ models, is that the original \citepalias{Meraxes} uses three free parameters for calculating the mass loading factor and energy coupling efficiency (given by Eqs.\ (14) and (13) in \citetalias{Meraxes}, respectively), whereas the modified version \citepalias{Qiu2019} only uses a single free parameter for both quantities (Eqs.\ (9) and (10) in \citetalias{Qiu2019}, respectively).
Not only does this reduce the number of parameters that need to be constrained by four (five if one also counts the \texttt{MergerBurstFactor}) compared to the previous emulators, but it also removes many correlation patterns between all the parameters.

Furthermore, the \citetalias{Qiu2019} \meraxes\ model uses UV \textit{luminosity function} (LF) and \textit{color-magnitude relation} (CMR) data from \citet{Bouwens2014,Bouwens2015} as observational constraints, not the SMF data.
To be able to handle this, \citetalias{Qiu2019} implemented three different parametrizations of the dust model proposed by \citet{Charlot00} in \meraxes.
We chose to use their \textit{dust-to-gas ratio} (DTG) parametrization in this work as it seemed to be the least constrained, giving the emulation process more freedom.
Despite this, the priors for all parameters as reported in \citetalias{Qiu2019} are still much more heavily constrained than the priors we used earlier (as given in \autoref{tab:M16_par}).

Finally, the \textit{initial mass function} (IMF) used in \citetalias{Qiu2019} \meraxes\ (Kroupa; \citealt{Kroupa_IMF}) is different from the one used in \citetalias{Meraxes} \meraxes\ (Salpeter; \citealt{Salpeter_IMF}).
Due to this change in the used IMF, the SMF data from \citet{Song2016} needs to be adjusted.
According to Eq.\ (2) in \citet{Speagle14}, the SMF masses can be shifted from a Salpeter IMF to a Kroupa IMF, by applying the following offset:
\begin{align}
    M_{*,K} &= 0.62M_{*,S},
\end{align}
with the $K$ and $S$ subscripts referring to the Kroupa and Salpeter IMFs, respectively.
This translates into an offset of $-0.21\,\mathrm{dex}$ for logarithmic masses.
For the emulators reported in this section, we have applied this offset to the SMF data from \citet{Song2016}.

Since the \citetalias{Qiu2019} \meraxes\ model uses different parameters than the original \citetalias{Meraxes} \meraxes, the parameter ranges and estimates given in \autoref{tab:M16_par} are no longer correct.
Therefore, we opted for using the same ranges and estimates for both the SMF and dust optical depth parameters, as described in Table 2 in \citetalias{Qiu2019}, which are given in \autoref{tab:Q19_par}.
For the data, we used the same data as used in \citetalias{Meraxes} and \citetalias{Qiu2019}, which is the observational data given by \citet{Song2016} (SMF), adjusted for a Kroupa IMF \citep{Kroupa_IMF}; and \citet{Bouwens2014,Bouwens2015} (LF/CMR), respectively.
The parameter ranges and estimates in \citetalias{Qiu2019} were calibrated manually by repeatedly using an MCMC estimator,\footnote{Y.\ Qiu, 2020, private communication, April 1st} which we kept in mind during the analysis.

\begin{table}
    \centering
    \begin{tabular}{|c|c|c|c|}
    \hline
        \textbf{Name} & \textbf{Min} & \textbf{Max} & \textbf{Estimate}\\
    \hline
        \texttt{SfCriticalSDNorm} & $0.001$ & $0.25$ & $0.01$ \\
        \texttt{SfEfficiency} & $0.05$ & $0.18$ & $0.1$ \\
        \texttt{SnEjectionEff} & $0.8$ & $2.2$ & $1.5$ \\
        \texttt{SnReheatEff} & $2.0$ & $15.0$ & $7.0$ \\
        \texttt{a} & $0.1$ & $0.65$ & $0.34$ \\
        \texttt{n} & $-2.5$ & $-0.8$ & $-1.6$ \\
        \texttt{s1} & $0.4$ & $2.2$ & $1.2$ \\
        \texttt{tauBC} & $0.0$ & $1000.0$ & $381.3$ \\
        \texttt{tauISM} & $0.0$ & $50.0$ & $13.5$ \\
    \hline
    \end{tabular}
    \caption{The parameters used by \prism\ for calculating the SMF and LF/CMR in \citetalias{Qiu2019} \meraxes, where all other (potentially important) parameters were set to their default values at all times.
    The boundaries and estimates were obtained from Table 2 in \citetalias{Qiu2019}.
    The names in the \textbf{first column} indicate the name that these parameters have in \citetalias{Qiu2019} \meraxes.}
    \label{tab:Q19_par}
\end{table}

\subsection{Comparing the two Meraxes models}
\label{subsec:Q19_SMF}
As the number of free parameters used for the mass loading factor and energy coupling efficiency in \citetalias{Qiu2019} \meraxes\ was reduced to one and the used IMF was changed, we first decided to make an emulator using solely SMF data.
We had concluded earlier that \citetalias{Meraxes} \meraxes\ was very difficult to constrain properly using only SMF data at $z = [7, 6, 5]$ and we wished to determine whether the parameter reduction and/or change in the IMF has any positive effects on the convergence behavior of the parameter space.
For this emulator, we assumed a model discrepancy variance \mdvar\ set to $(z_i/100)^2$, with $z_i$ being the value of the corresponding observational data point.
This results in values that are very similar to the static $10^{-4}$ we used earlier (as the SMF data values are mapped to an $\arctan$ function and thus of the order of unity), but makes them dependent on the data value.
We did this in order to prepare for the LF/CMR data points we will use later, as their values are of the order of $10^{-4}$ and thus such a model discrepancy variance would not work well for them.
The statistics of this emulator are shown in \autoref{tab:Q19_SMF_stats}.

\begin{table}
    \centering
    \begin{tabular}{|c|r|l|c|l|}
    \hline
        \multicolumn{1}{|c|}{$i_{\mathrm{emul}}$} &
        \multicolumn{1}{|c|}{$n_{\mathrm{eval}}$} & 
        \multicolumn{1}{|c|}{$I_{\mathrm{cut,n}}$} & 
        \multicolumn{1}{|c|}{$n_{\mathrm{wild}}$} &
        \multicolumn{1}{|c|}{$f_{\mathrm{space}}$} \\
    \hline
        $1$ & $500$ & $[4.0, 3.5, 3.2, 3.0]$ & $0$ & $19.6\%$ \\
        $2$ & $941$ & $[3.0, 2.5, 2.2, 2.0]$ & $0$ & $0.203\%$ \\
        $3$ & $1,191$ & $[2.8, 2.5, 2.3]$ & $0$ & $0.0575\%$ \\
    \hline
    \end{tabular}
    \caption{Statistics for the SMF-only \citetalias{Qiu2019} \meraxes\ emulator.
    The \textbf{first column} specifies the emulator iteration $i_{\mathrm{emul}}$ this row is about.
    The \textbf{next three columns} provide the number of model evaluations $n_{\mathrm{eval}}$; the non-wildcard implausibility cut-offs $I_{\mathrm{cut,n}}$; and the number of implausibility wildcards $n_{\mathrm{wild}}$ used for this emulator iteration.
    Finally, the \textbf{last column} gives the fraction of parameter space remaining $f_{\mathrm{space}}$ after this emulator iteration was analyzed.}
    \label{tab:Q19_SMF_stats}
\end{table}

Looking at \autoref{tab:Q19_SMF_stats}, we can see that the parameters in this emulator are constrained much more heavily than in previous emulators.
As a reminder, every implausibility wildcard causes the next highest implausibility measure to be ignored when evaluating the emulator.
Despite using no wildcards at all, this emulator still had $19.6\%$ of parameter space remaining in iteration $1$.
This may seem like much compared to the statistics of previous emulators (that do use wildcards), but this can be easily explained by noting that the parameter ranges are far smaller than those used before.
Therefore only a potentially more interesting/plausible space was available to begin with.

We also note that there is a large reduction in plausible space between iterations $1$ and $2$, corresponding to a factor of about $100$.
Given that the implausibility cut-offs $I_{\mathrm{cut,n}}$ were not reduced by a large amount when going to iteration $2$ and neither had any implausibility wildcards involved, this implies that many comparison data points were just barely plausible in iteration $1$.
This would mean that a majority of the data points have a high constraining power in \citetalias{Qiu2019} \meraxes\ and similar implausibility values.
The projection figures will allow us to see whether this is the case.

\begin{figure*}
    \centering
    \subfloat{\includegraphics[width=0.24\textwidth]{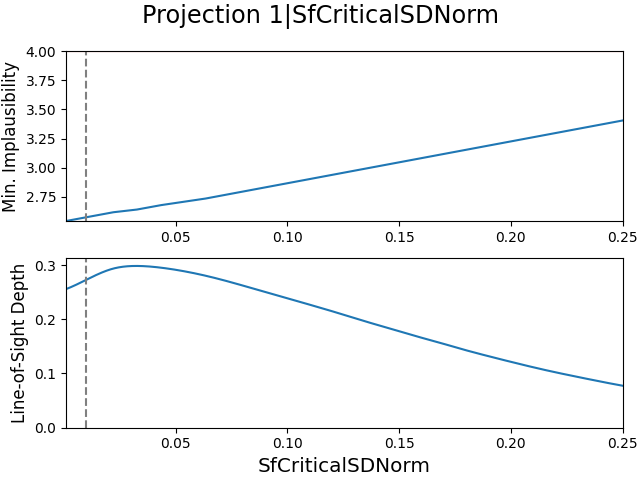}}
    \subfloat{\includegraphics[width=0.24\textwidth]{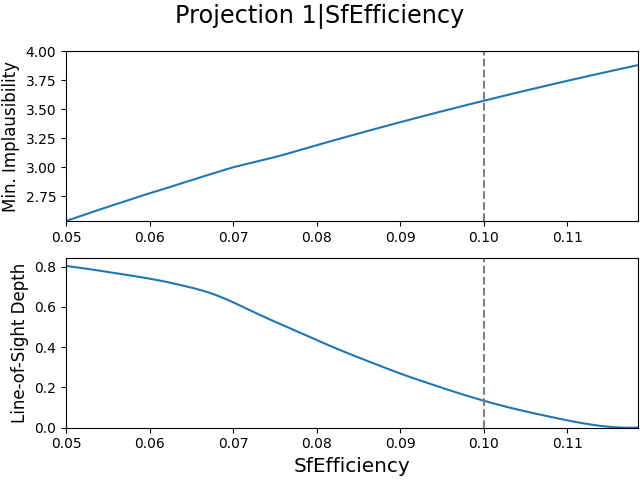}}
    \subfloat{\includegraphics[width=0.24\textwidth]{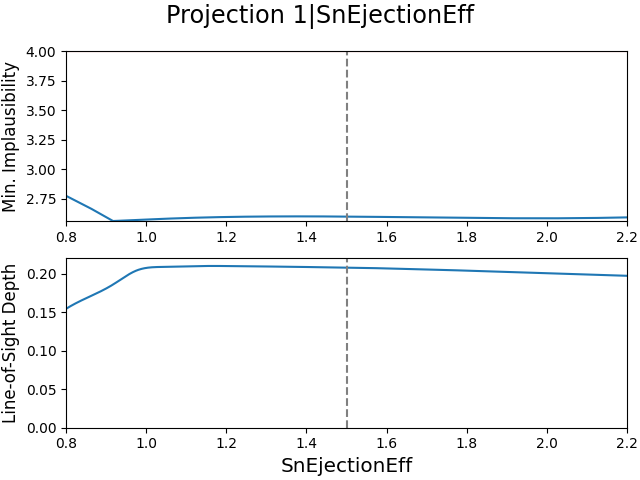}}
    \subfloat{\includegraphics[width=0.24\textwidth]{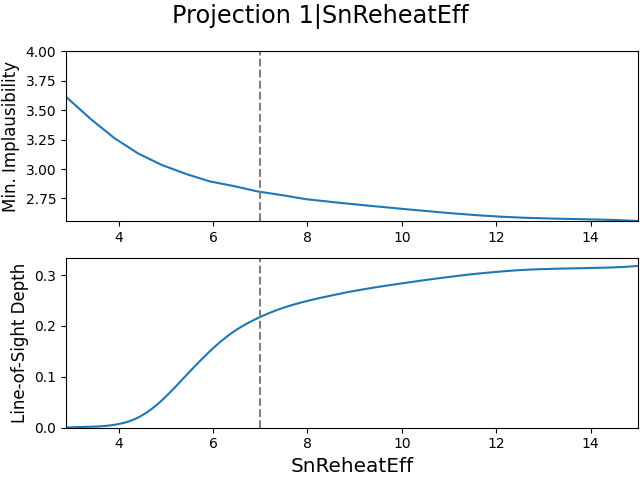}} \\
    \subfloat{\includegraphics[width=0.24\textwidth]{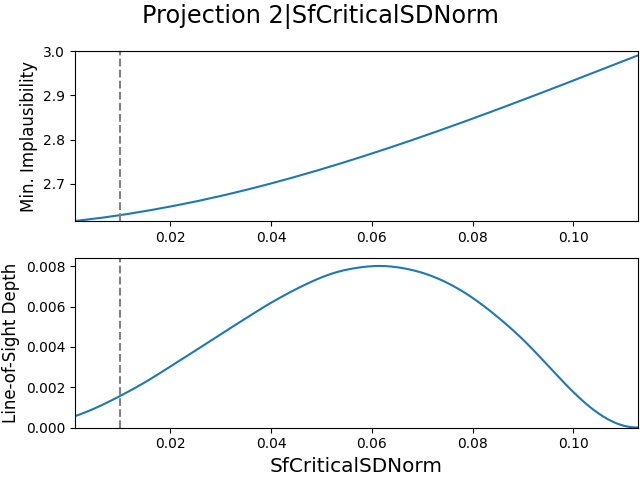}}
    \subfloat{\includegraphics[width=0.24\textwidth]{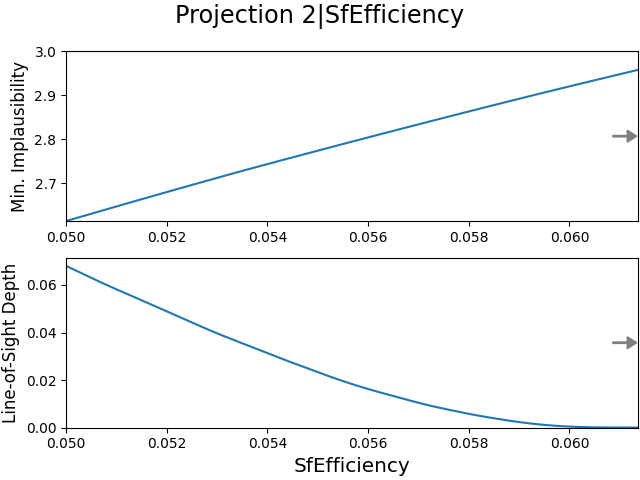}}
    \subfloat{\includegraphics[width=0.24\textwidth]{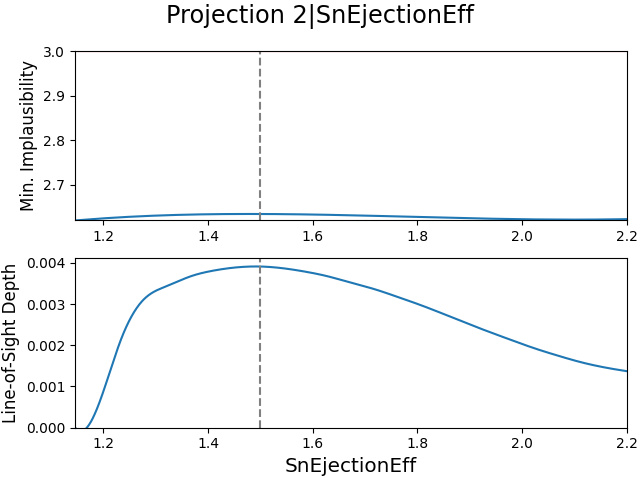}}
    \subfloat{\includegraphics[width=0.24\textwidth]{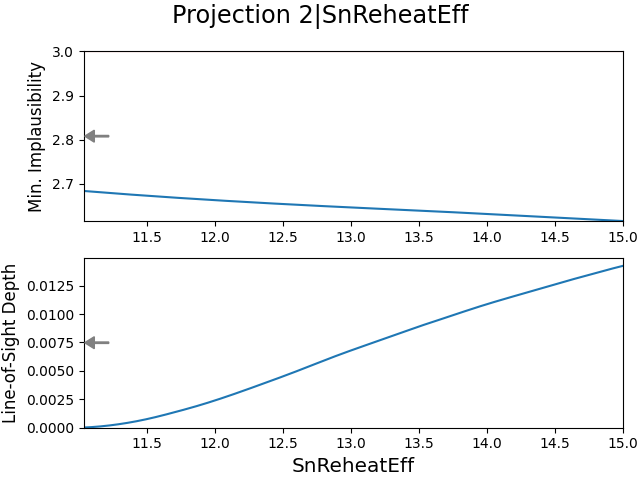}} \\
    \subfloat{\includegraphics[width=0.24\textwidth]{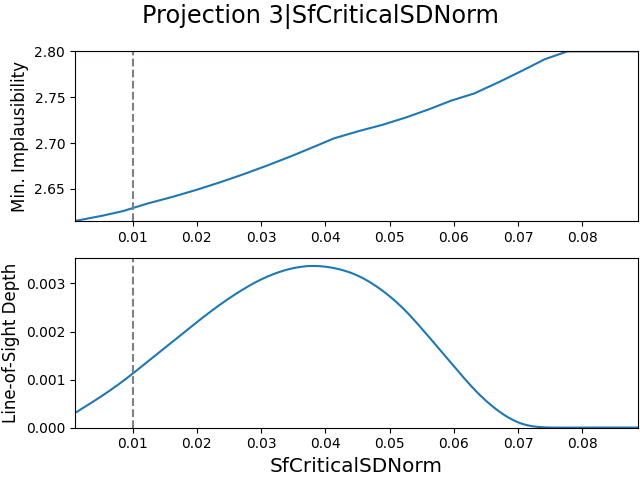}}
    \subfloat{\includegraphics[width=0.24\textwidth]{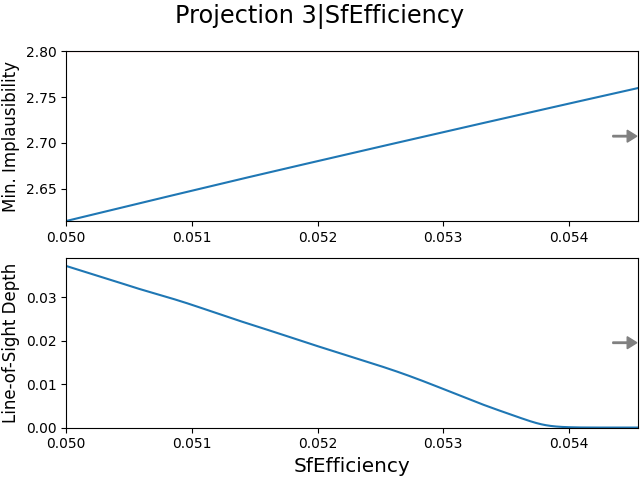}}
    \subfloat{\includegraphics[width=0.24\textwidth]{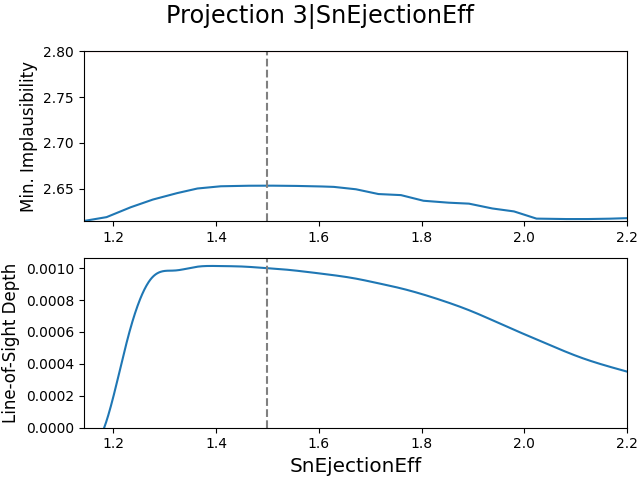}}
    \subfloat{\includegraphics[width=0.24\textwidth]{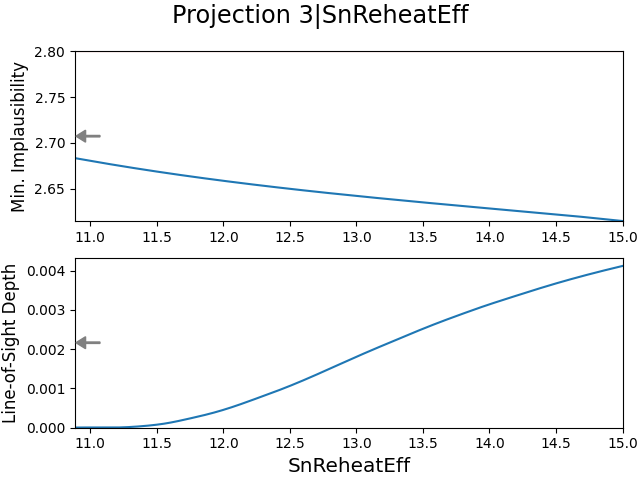}} \\
    \subfloat{\includegraphics[width=0.24\textwidth]{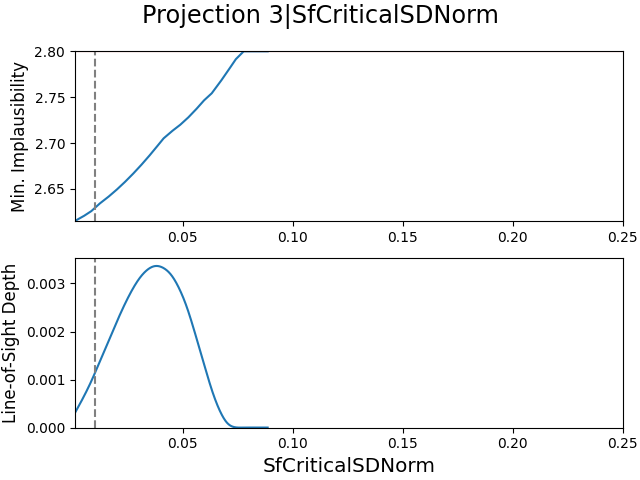}}
    \subfloat{\includegraphics[width=0.24\textwidth]{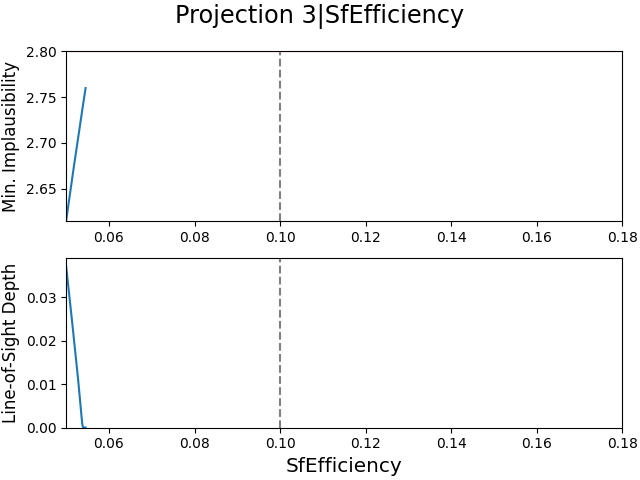}}
    \subfloat{\includegraphics[width=0.24\textwidth]{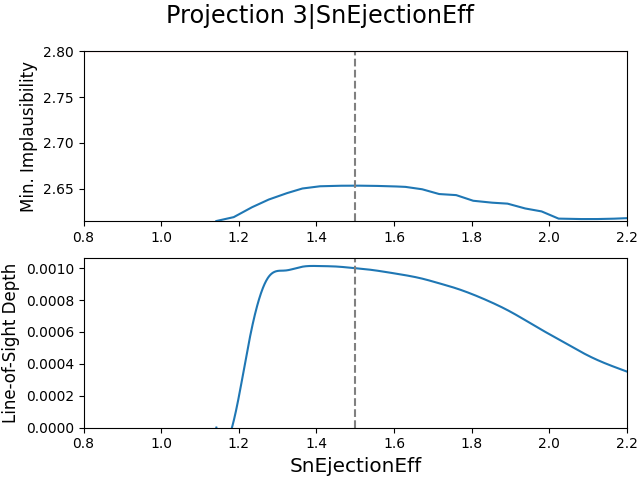}}
    \subfloat{\includegraphics[width=0.24\textwidth]{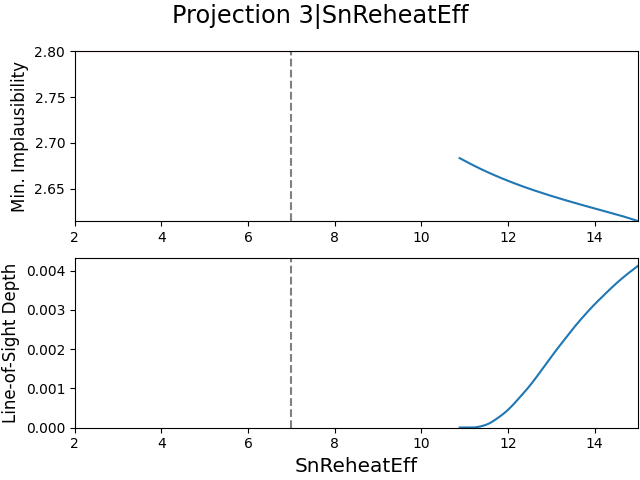}}
    \caption{2D projection figures of the SMF-only \citetalias{Qiu2019} \meraxes\ emulator at all three iterations, showing the four parameters related to the star formation and supernova feedback.
    The \textbf{dashed lines} show the estimated value of the corresponding parameter as given in \autoref{tab:Q19_par}.
    In case the parameter estimate is outside of the plotted value range, an \textbf{arrow} pointing in the direction of the estimate is shown instead.
    \textbf{First row:} $i_{\mathrm{emul}}=1$.
    \textbf{Second row:} $i_{\mathrm{emul}}=2$.
    \textbf{Bottom rows:} $i_{\mathrm{emul}}=3$ with the \textbf{final row} showing the full parameter range instead of only the defined range, but it is otherwise equivalent to the \textbf{third row}.}
    \label{fig:Q19_SMF_results_2D}
\end{figure*}

The 2D projections of the parameters related to the star formation and supernova feedback processes in \meraxes, are shown in \autoref{fig:Q19_SMF_results_2D} for all three emulator iterations.
The dashed lines in the figures show the estimated value of the corresponding parameter as given in \autoref{tab:Q19_par}.
Note that the parameter ranges in the projection figures differ between emulator iterations, as the emulator was only defined over that specific range.
For comparison, we show the full parameter range for the parameters in emulator iteration $3$ in the final row. 

Using this information, we can see some interesting trends in the projection figures.
For example, all parameters except for the energy coupling efficiency, \texttt{SnEjectionEff}, seem to be moving away from their estimate.
This could either mean that the parameter estimates are potentially biased, as we encountered earlier with the estimates given in \autoref{tab:M16_par}, or that the LF/CMR data constrains these parameters differently.
Despite this however, the parameters are much more well-behaved compared to the projections in \autoref{fig:M16_mock_results} and \autoref{fig:M16_SMF_results}, confirming that the parameter reduction has a positive effect on the convergence behavior.

Furthermore, we can see that our observation from earlier is indeed correct, that most data points have similar implausibility values.
This is particularly evident in the projection figure of the energy coupling efficiency, \texttt{SnEjectionEff}, at iterations $2$ and $3$ (and to a lesser extent, the mass loading factor, \texttt{SnReheatEff}, shows this as well).
As the y-axis in the top subplot ranges from the lowest (reachable) implausibility value to the highest (plausible) implausibility value, it shows that there is little variation in the values it can take.
Taking into account the implausibility cut-offs reported in \autoref{tab:Q19_SMF_stats}, this behavior can only be obtained when most data points have similar implausibility values and thus similar constraining power as well.

\begin{figure*}
    \centering
    \subfloat{\includegraphics[width=0.24\textwidth]{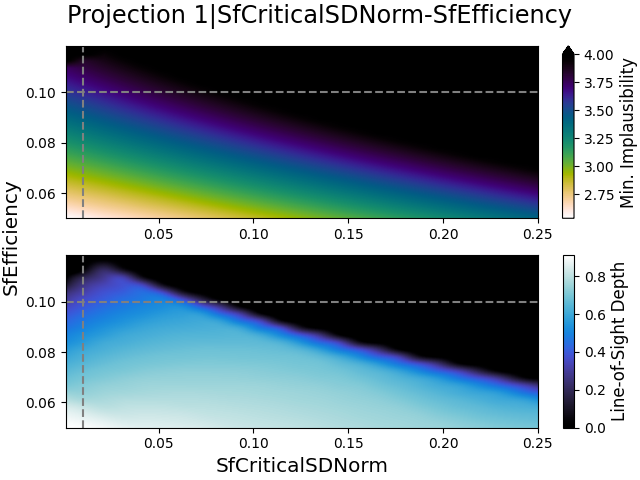}}
    \subfloat{\includegraphics[width=0.24\textwidth]{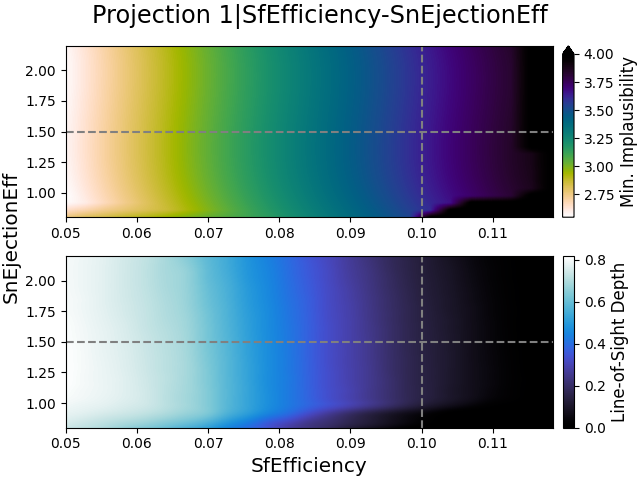}}
    \subfloat{\includegraphics[width=0.24\textwidth]{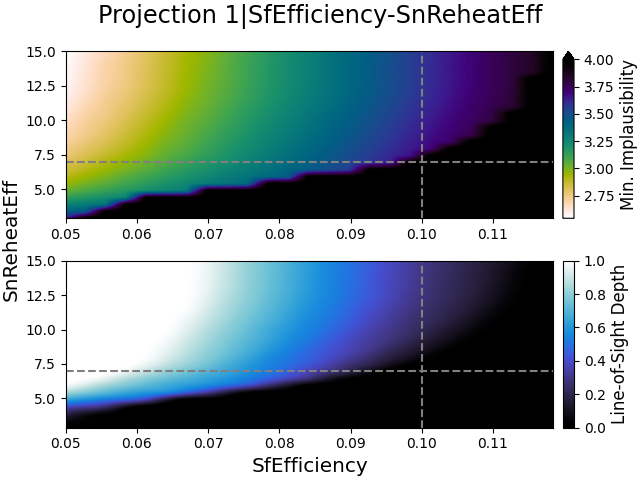}}
    \subfloat{\includegraphics[width=0.24\textwidth]{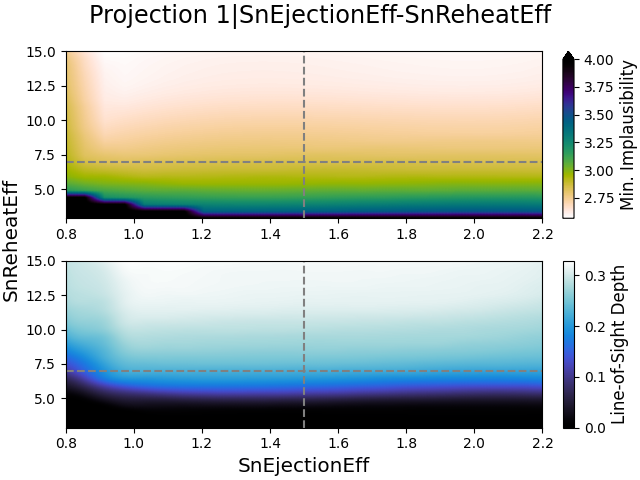}} \\
    \subfloat{\includegraphics[width=0.24\textwidth]{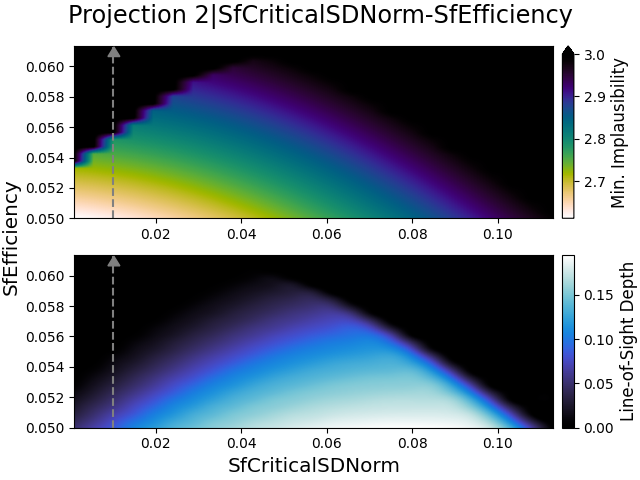}}
    \subfloat{\includegraphics[width=0.24\textwidth]{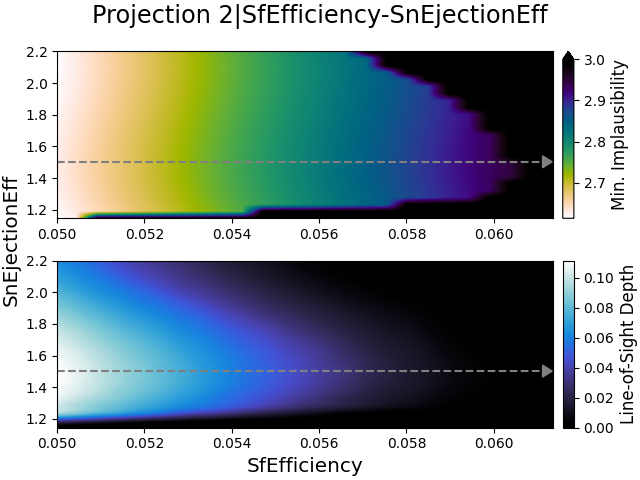}}
    \subfloat{\includegraphics[width=0.24\textwidth]{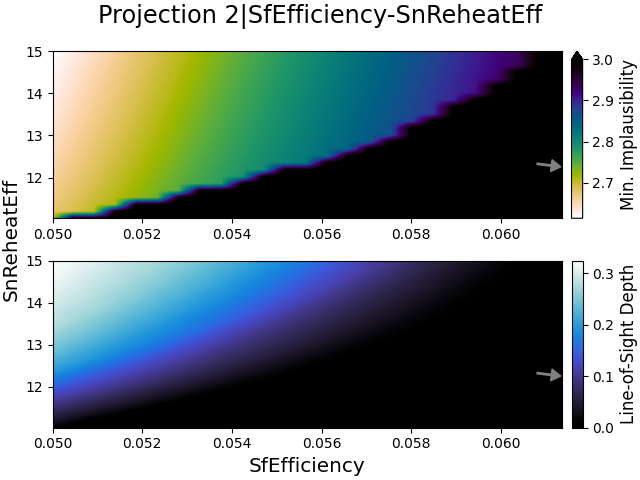}}
    \subfloat{\includegraphics[width=0.24\textwidth]{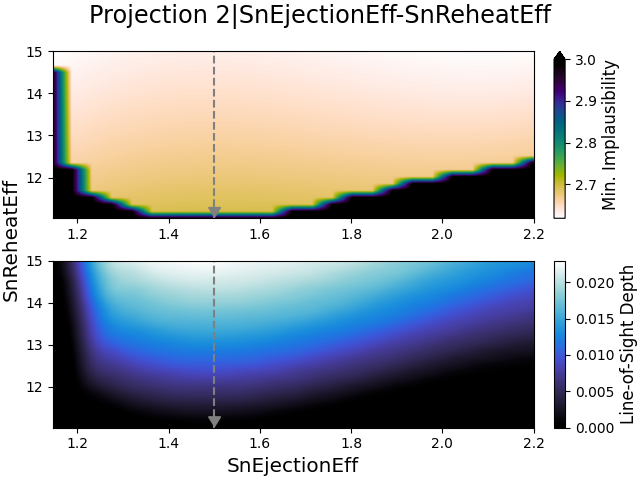}} \\
    \subfloat{\includegraphics[width=0.24\textwidth]{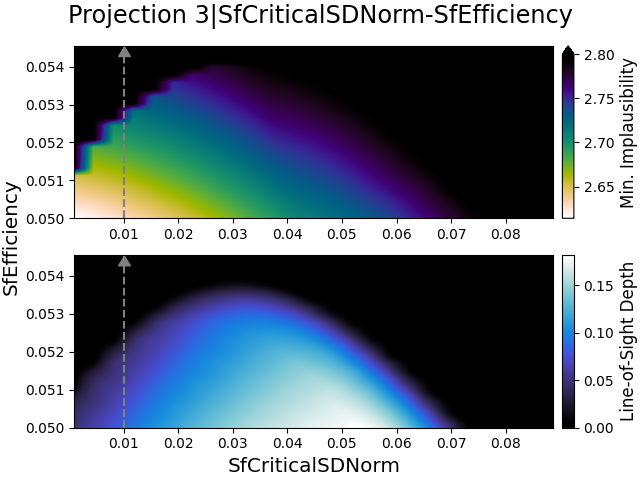}}
    \subfloat{\includegraphics[width=0.24\textwidth]{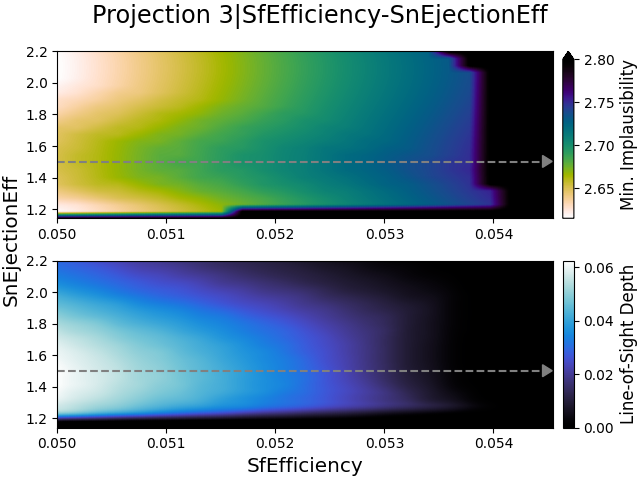}}
    \subfloat{\includegraphics[width=0.24\textwidth]{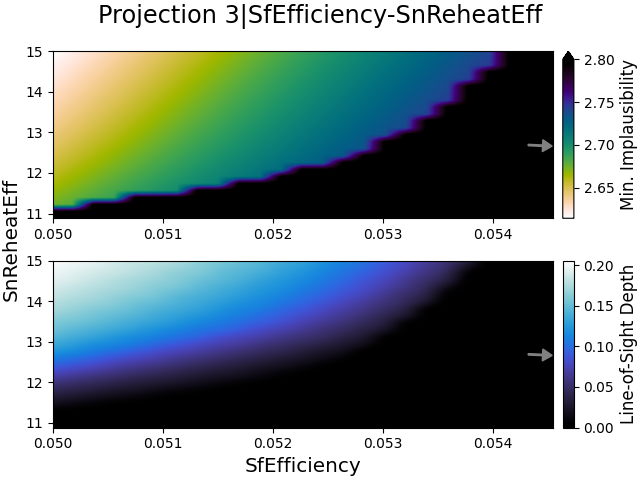}}
    \subfloat{\includegraphics[width=0.24\textwidth]{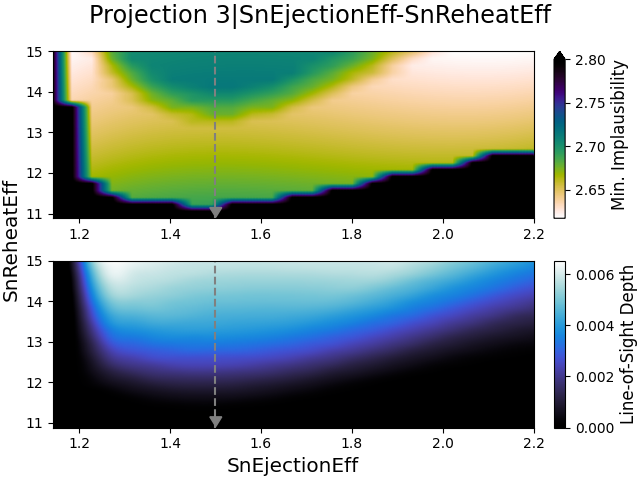}} \\
    \subfloat{\includegraphics[width=0.24\textwidth]{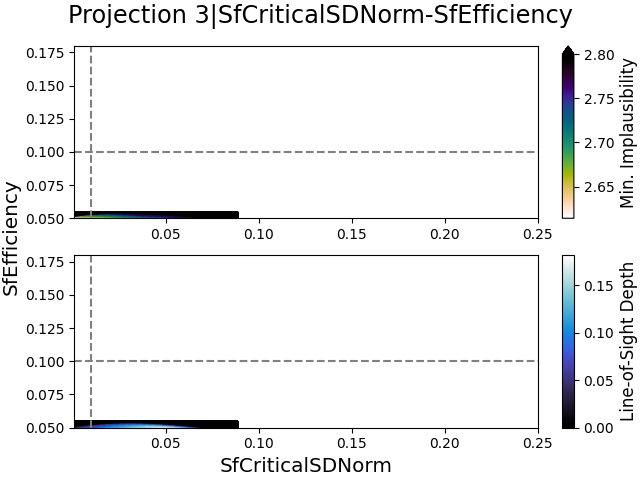}}
    \subfloat{\includegraphics[width=0.24\textwidth]{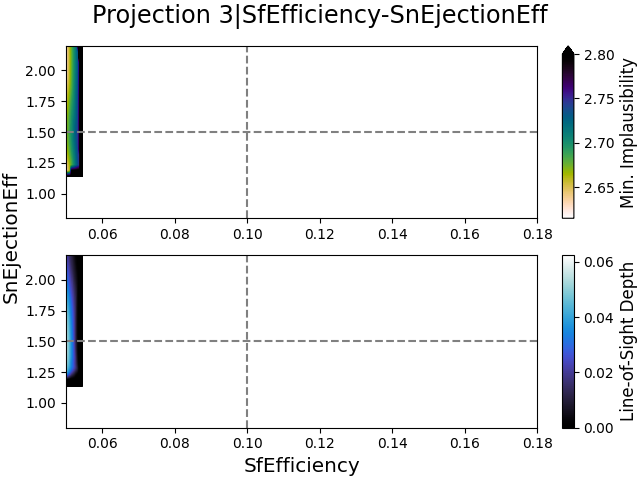}}
    \subfloat{\includegraphics[width=0.24\textwidth]{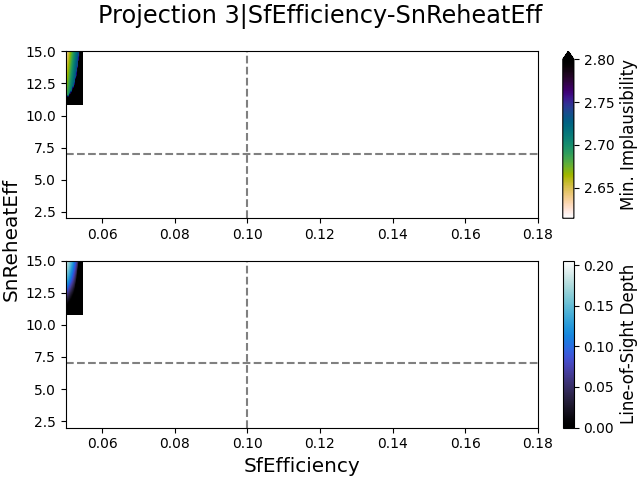}}
    \subfloat{\includegraphics[width=0.24\textwidth]{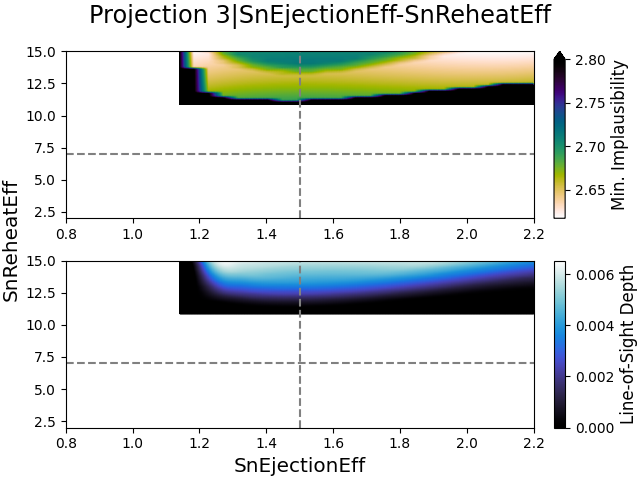}}
    \caption{3D projection figures of the SMF-only \citetalias{Qiu2019} \meraxes\ emulator at all three iterations, showing the four parameters related to the star formation and supernova feedback.
    The \textbf{first three columns} show the correlation between the star formation efficiency \texttt{SfEfficiency} and the other parameters.
    The \textbf{last column} shows the correlation between the energy coupling efficiency \texttt{SnEjectionEff} and the mass loading factor \texttt{SnReheatEff}.
    The \textbf{dashed lines} show the estimated value of the corresponding parameter as given in \autoref{tab:Q19_par}.
    In case either parameter estimate is outside of the plotted value range, an \textbf{arrow} pointing in the direction of the intersection of the estimates is shown instead.
    \textbf{First row:} $i_{\mathrm{emul}}=1$.
    \textbf{Second row:} $i_{\mathrm{emul}}=2$.
    \textbf{Bottom rows:} $i_{\mathrm{emul}}=3$ with the \textbf{final row} showing the full parameter range instead of only the defined range, but it is otherwise equivalent to the \textbf{third row}.}
    \label{fig:Q19_SMF_results_3D}
\end{figure*}

In \autoref{fig:Q19_SMF_results_3D}, we show the 3D projections for the parameters related to the star formation and supernova feedback processes, for all three emulator iterations.
Because the star formation efficiency \texttt{SfEfficiency} is commonly the most well constrained and most correlated with the other parameters, the first three projections show the correlations between \texttt{SfEfficiency} and the other parameters.
As the energy coupling efficiency \texttt{SnEjectionEff} and the mass loading factor \texttt{SnReheatEff} should be heavily correlated with each other, in addition to being modified in \citetalias{Qiu2019} \meraxes, we show their 3D projections as well.

Looking at these projections, we can see that the star formation efficiency is indeed heavily correlated with the other parameters, revealing strong relations in all projection figures.
Despite this however, it appears that the energy coupling efficiency, \texttt{SnEjectionEff}, cannot be effectively constrained.
This is evident in both projection figures that show this parameter, which can be found in the second and fourth columns of \autoref{fig:Q19_SMF_results_3D}.
In fact, it appears that nearly all values for this parameter except its estimate are equally plausible, which is particularly visible in the iteration $3$ figure in the fourth column.
This strongly implies that the SMF data cannot constrain this parameter, which might become important later.

Finally, we note that all parameters, with maybe the exception of the energy coupling efficiency \texttt{SnEjectionEff}, seem to be limited by their own priors, implying that their best values lie outside of their parameter ranges.
This is particularly clear in their 3D projection figures in the third column of \autoref{fig:Q19_SMF_results_3D}.
This is another indication that the parameter priors might be too constrained, or that the LF/CMR data constrains these parameters differently.
To test this hypothesis, we checked what the SMF looks like when the best plausible samples found in the emulator are used.

\begin{figure*}
    \centering
    \includegraphics[width=\textwidth]{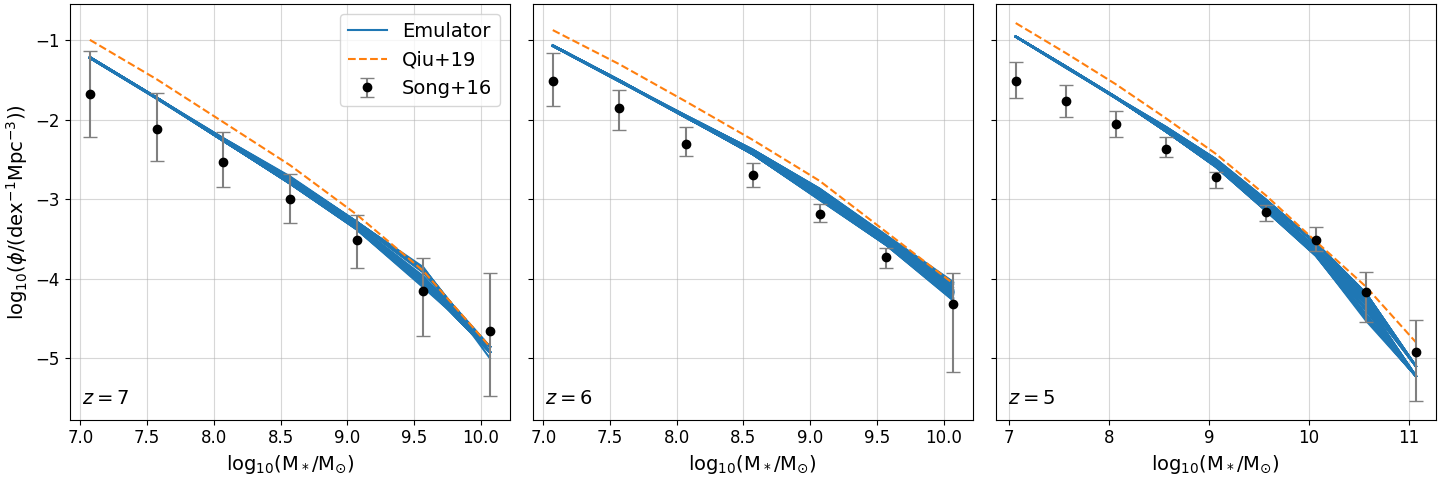}
    \caption{Realizations of the stellar mass functions at redshifts $z=[7, 6, 5]$ using the results of the SMF-only \citetalias{Qiu2019} \meraxes\ emulator.
    All realizations were created by directly evaluating \citetalias{Qiu2019} \meraxes.
    The \textbf{solid lines} use the $50$ best plausible samples out of $2,500$ in the emulator at iteration $3$.
    The \textbf{dashed line} uses the parameter estimates as given in \autoref{tab:Q19_par}.
    The \textbf{dots} show the SMF data from \citet{Song2016} adjusted for a Kroupa IMF, with corresponding standard deviations.}
    \label{fig:Q19_SMF_fits}
\end{figure*}

In \autoref{fig:Q19_SMF_fits}, we show a comparison between the SMFs created by $50$ plausible samples in the emulator and the SMF given by the parameter estimates from \autoref{tab:Q19_par}.
Because an emulator consists of approximations, it is unable to provide a true best parameter fit, and therefore we evaluated the emulator $2,500$ times within plausible space at iteration $3$, and selected the $50$ best plausible samples.
These $50$ samples were then evaluated in \citetalias{Qiu2019} \meraxes\ to produce the solid lines shown in the figure.

From this figure, we can see that the emulator samples (solid lines) apparently fit the data much better than the parameter estimates from \autoref{tab:Q19_par} (dashed line).
However, even though the fits are better, the emulator still appears to be overestimating the number density of galaxies at low masses, for all redshifts.
If we take a look at \autoref{fig:Q19_SMF_results_2D}, we can see that the lowest implausibility values are reached at either boundary for all parameters.
This means that these $50$ plausible samples all have values that are at the minimum/maximum value they can take.
Therefore, this figure confirms the hypothesis we made earlier, that the parameter priors are too constrained, and/or that the LF/CMR data constrains these parameters differently.
In order to test which of the two statements is true, we explore whether the trend is still apparent when using the LF/CMR data in the next section.

\subsection{Exploring Meraxes using both SMF and luminosity data}
\label{subsec:Q19_full}
Now that we have analyzed the \citetalias{Qiu2019} \meraxes\ model and studied how it affects the behavior of the SMF parameters, it is time to analyze the model using all nine parameters.
As mentioned before, we use the same data as used in \citetalias{Meraxes} and \citetalias{Qiu2019} (i.e., \citealt{Bouwens2014,Bouwens2015,Song2016}), and the same parameter ranges as given in \autoref{tab:Q19_par}.
To allow for a better comparison with the previous emulator, this emulator also uses a model discrepancy variance \mdvar\ of $(z_i/100)^2$.
The statistics of this emulator are shown in \autoref{tab:Q19_full_stats}.

\begin{table}
    \centering
    \begin{tabular}{|c|r|l|c|l|}
    \hline
        \multicolumn{1}{|c|}{$i_{\mathrm{emul}}$} &
        \multicolumn{1}{|c|}{$n_{\mathrm{eval}}$} & 
        \multicolumn{1}{|c|}{$I_{\mathrm{cut,n}}$} & 
        \multicolumn{1}{|c|}{$n_{\mathrm{wild}}$} &
        \multicolumn{1}{|c|}{$f_{\mathrm{space}}$} \\
    \hline
        $1$ & $500$ & $[4.0, 3.5, 3.2, 3.0]$ & $3$ & $9.89\%$ \\
        $2$ & $1,068$ & $[4.0, 3.5, 3.2, 3.0]$ & $1$ & $1.39\%$ \\
        $3$ & $1,519$ & $[4.0, 3.5, 3.2, 3.0]$ & $0$ & $0.226\%$ \\
        $4$ & $1,481$ & $[3.5, 3.0, 2.7, 2.5]$ & $0$ & $0.00489\%$ \\
    \hline
    \end{tabular}
    \caption{Statistics for the full \citetalias{Qiu2019} \meraxes\ emulator.
    The \textbf{first column} specifies the emulator iteration $i_{\mathrm{emul}}$ this row is about.
    The \textbf{next three columns} provide the number of model evaluations $n_{\mathrm{eval}}$; the non-wildcard implausibility cut-offs $I_{\mathrm{cut,n}}$; and the number of implausibility wildcards $n_{\mathrm{wild}}$ used for this emulator iteration.
    Finally, the \textbf{last column} gives the fraction of parameter space remaining $f_{\mathrm{space}}$ after this emulator iteration was analyzed.}
    \label{tab:Q19_full_stats}
\end{table}

As this emulator uses over three times the number of data points as the previous one ($74$ vs.\ $24$) and because we expect the LF/CMR data to constrain the model differently, we intentionally converged the emulator in a slower, more conservative fashion.
We therefore have three implausibility wildcards for the first iteration.
This can potentially also provide us with more information on the value of using the LF/CMR data as additional constraints for \meraxes.

Looking at \autoref{tab:Q19_full_stats}, we can see that the emulator converged rather smoothly throughout all iterations.
However, we do note that there is a reduction of roughly a factor $50$ in plausible space between iterations $3$ and $4$.
Given that the difference in the implausibility parameters between these two iterations is not very large, this is implies that all data points have similar implausibility values and thus similar constraining power.
As we found a similar trend for the SMF-only emulator (see \autoref{tab:Q19_SMF_stats}), we will investigate later using projection figures and emulator evaluations whether this is the case.

\begin{figure*}[htb!]
    \centering
    \subfloat{\includegraphics[width=0.24\textwidth]{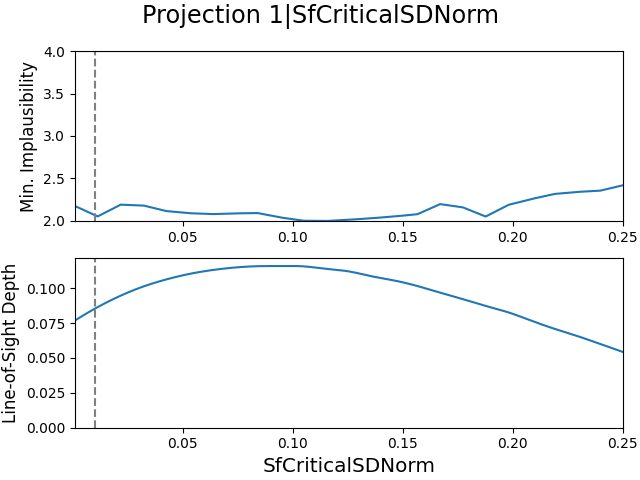}}
    \subfloat{\includegraphics[width=0.24\textwidth]{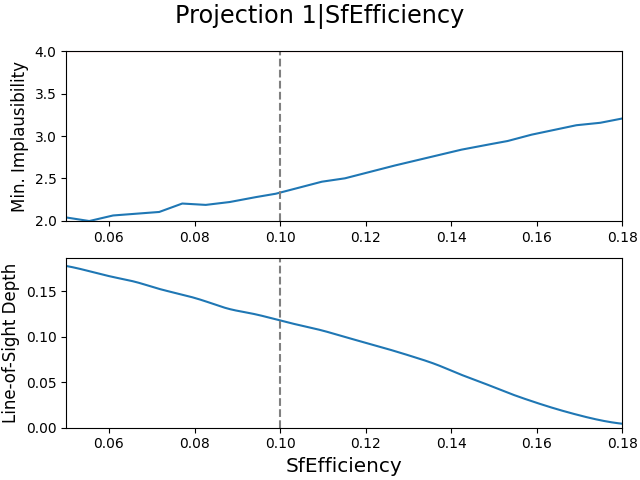}}
    \subfloat{\includegraphics[width=0.24\textwidth]{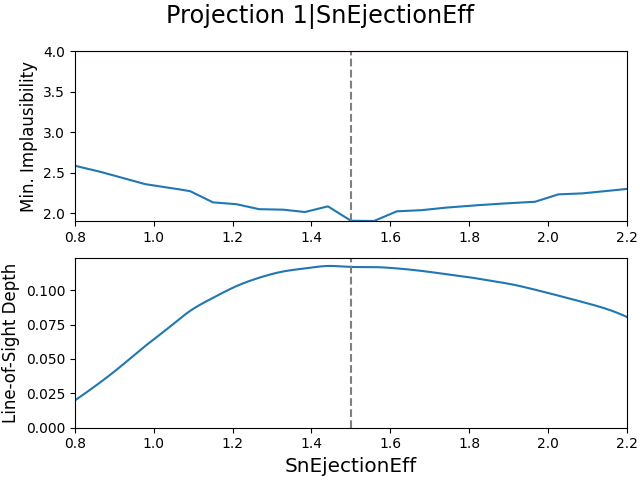}}
    \subfloat{\includegraphics[width=0.24\textwidth]{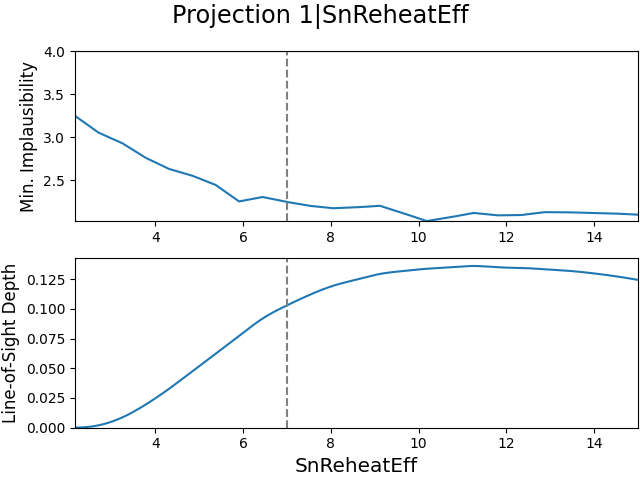}} \\
    \subfloat{\includegraphics[width=0.24\textwidth]{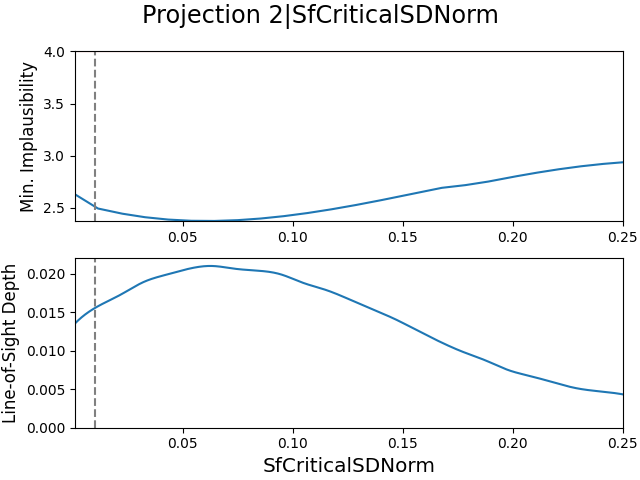}}
    \subfloat{\includegraphics[width=0.24\textwidth]{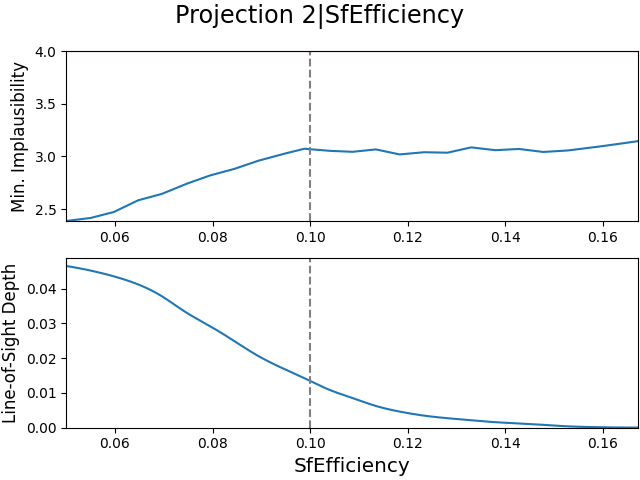}}
    \subfloat{\includegraphics[width=0.24\textwidth]{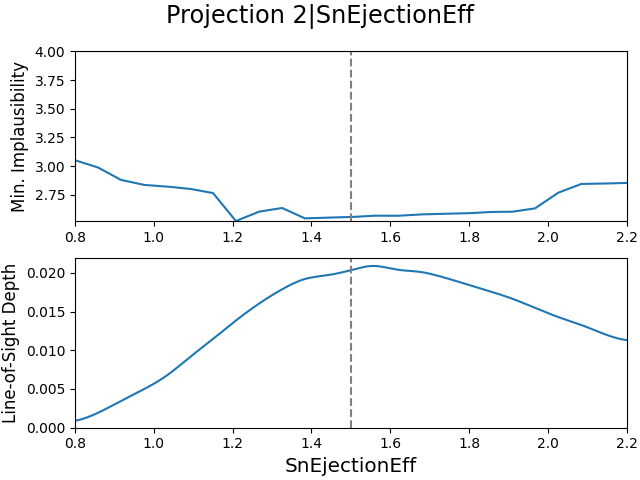}}
    \subfloat{\includegraphics[width=0.24\textwidth]{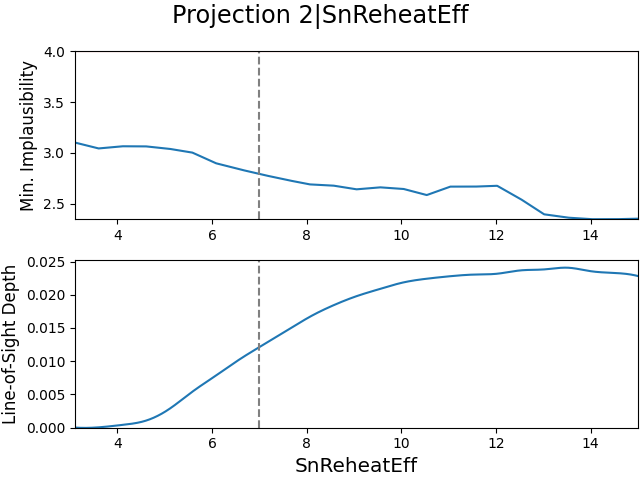}} \\
    \subfloat{\includegraphics[width=0.24\textwidth]{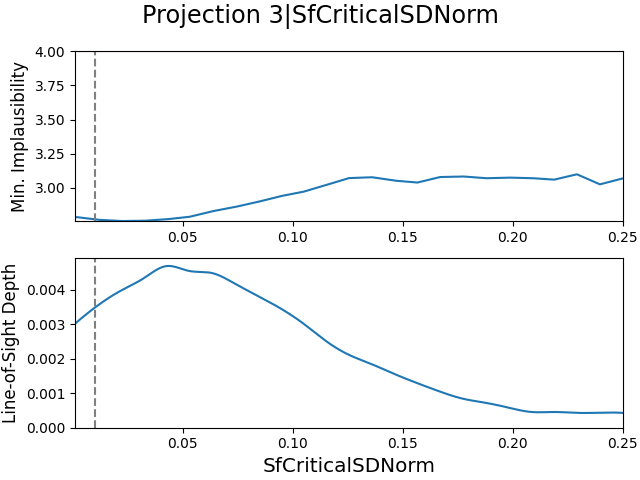}}
    \subfloat{\includegraphics[width=0.24\textwidth]{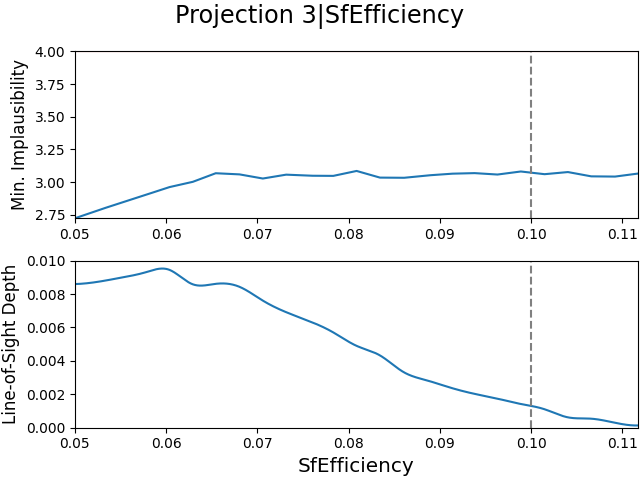}}
    \subfloat{\includegraphics[width=0.24\textwidth]{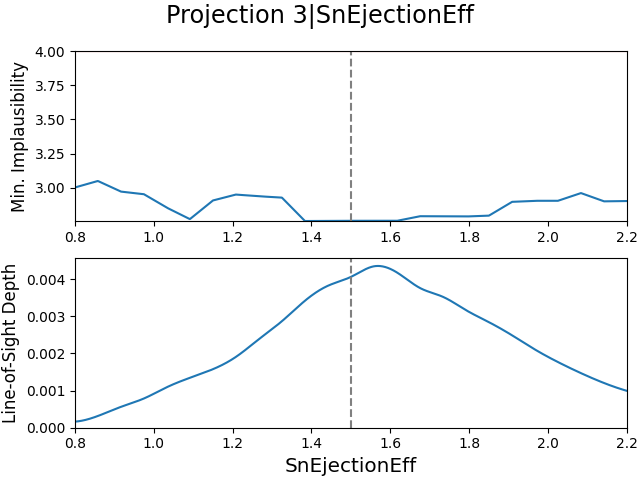}}
    \subfloat{\includegraphics[width=0.24\textwidth]{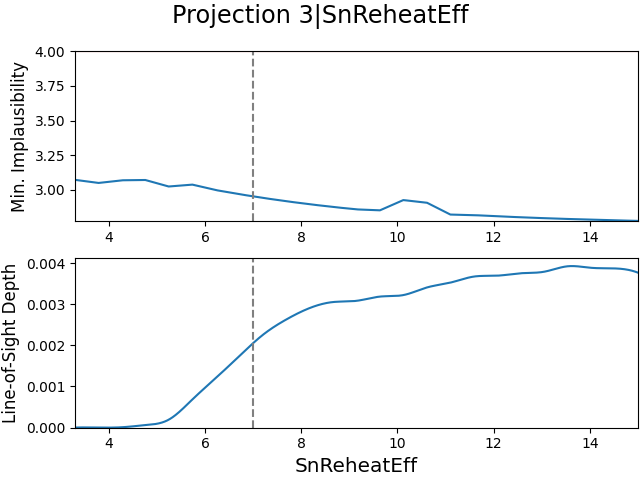}} \\
    \subfloat{\includegraphics[width=0.24\textwidth]{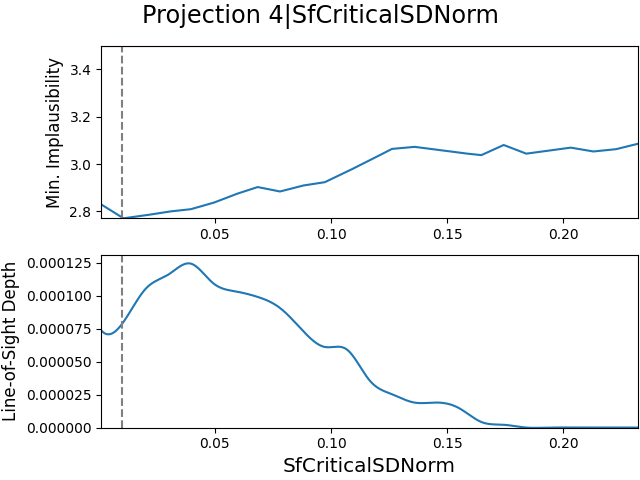}}
    \subfloat{\includegraphics[width=0.24\textwidth]{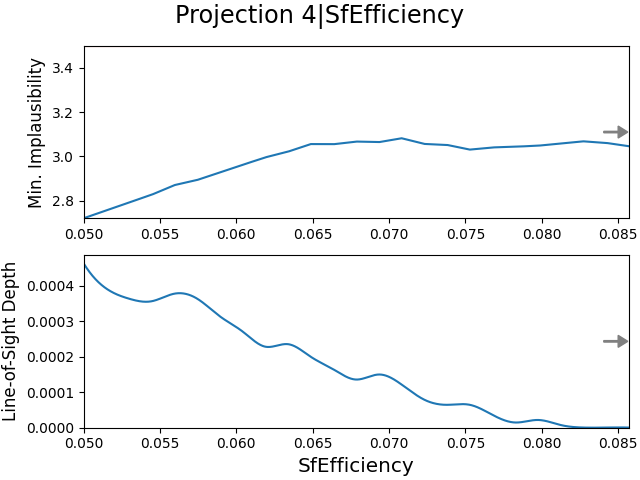}}
    \subfloat{\includegraphics[width=0.24\textwidth]{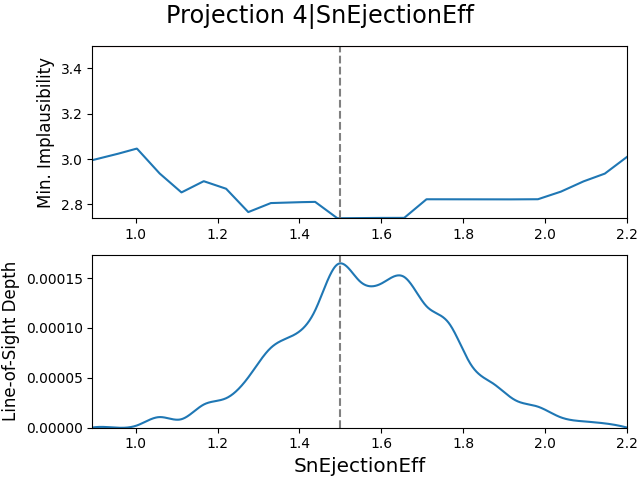}}
    \subfloat{\includegraphics[width=0.24\textwidth]{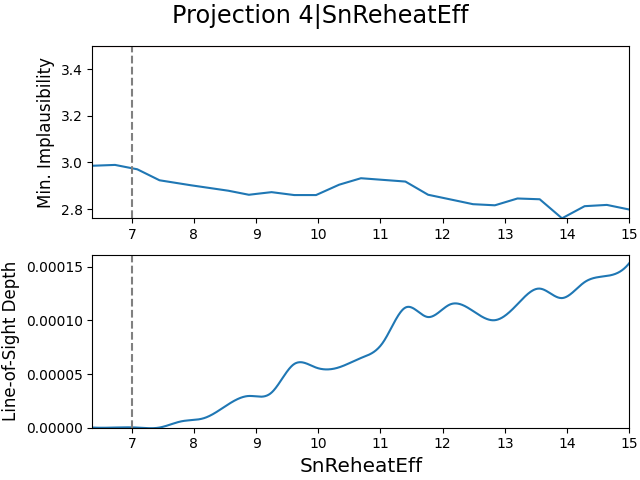}} \\
    \subfloat{\includegraphics[width=0.24\textwidth]{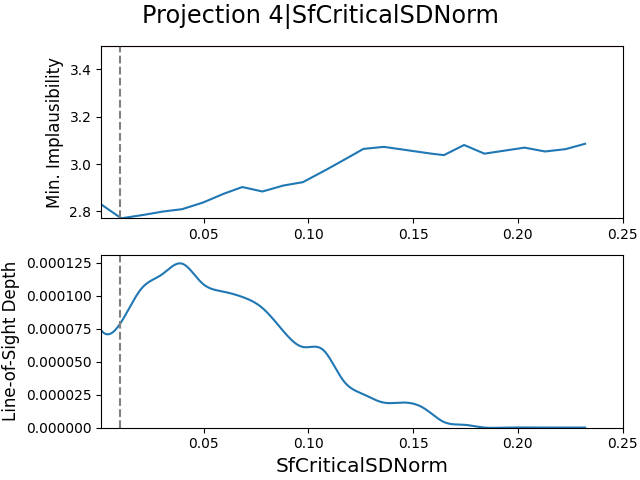}}
    \subfloat{\includegraphics[width=0.24\textwidth]{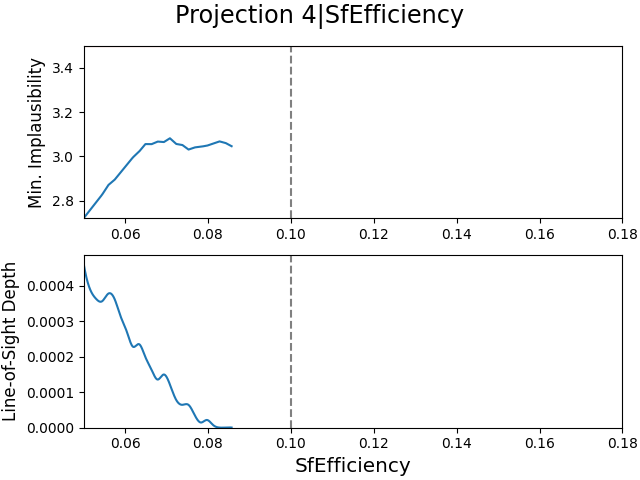}}
    \subfloat{\includegraphics[width=0.24\textwidth]{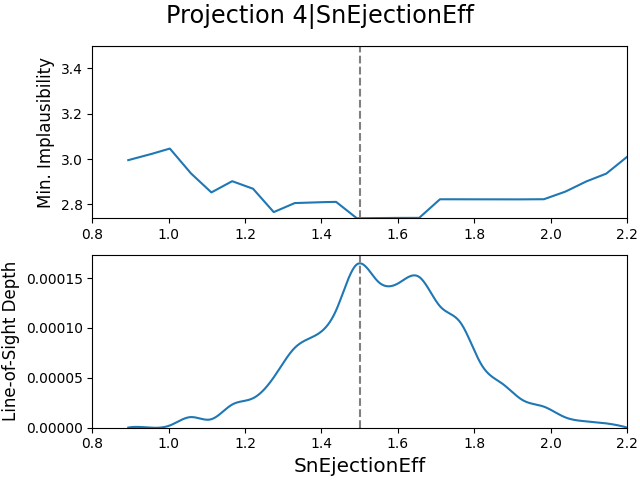}}
    \subfloat{\includegraphics[width=0.24\textwidth]{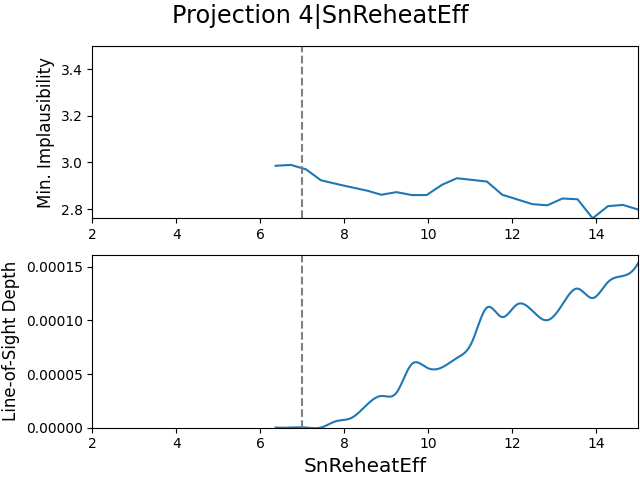}}
    \caption{2D projection figures of the full \citetalias{Qiu2019} \meraxes\ emulator at all four iterations, showing the four parameters related to the star formation and supernova feedback.
    The \textbf{dashed lines} show the estimated value of the corresponding parameter as given in \autoref{tab:Q19_par}.
    In case the parameter estimate is outside of the plotted value range, an \textbf{arrow} pointing in the direction of the estimate is shown instead.
    \textbf{First row:} $i_{\mathrm{emul}}=1$.
    \textbf{Second row:} $i_{\mathrm{emul}}=2$.
    \textbf{Third row:} $i_{\mathrm{emul}}=3$.
    \textbf{Bottom rows:} $i_{\mathrm{emul}}=4$ with the \textbf{final row} showing the full parameter range instead of only the defined range, but it is otherwise equivalent to the \textbf{fourth row}.}
    \label{fig:Q19_full_results_SMF_2D}
\end{figure*}

The 2D projections of the parameters related to the star formation and supernova feedback processes in \meraxes\ are shown in \autoref{fig:Q19_full_results_SMF_2D} for all four emulator iterations.
The dashed lines in the figures show the estimated value of the corresponding parameter as given in \autoref{tab:Q19_par}.
As with the previous emulator, the parameter ranges differ between emulator iterations.

First of all, we look at the top row in \autoref{fig:Q19_full_results_SMF_2D}, which shows the 2D projection figures for emulator iteration $1$.
Given that this iteration still had $9.89\%$ of parameter space remaining (according to \autoref{tab:Q19_full_stats}), we do not expect to see much structure here.
However, we can see that there is a reasonable amount of structure in them, which shows us that the parameters agree relatively well with their estimates.

In iteration $2$ however, we note that this trend is starting to change for the star formation efficiency, \texttt{SfEfficiency}, and mass loading factor, \texttt{SnReheatEff}, which move away from their estimates.
This becomes increasingly apparent over the next two iterations.
When we compare the iteration $4$ projections in \autoref{fig:Q19_full_results_SMF_2D} to the iteration $3$ projections in \autoref{fig:Q19_SMF_results_2D}, we can see that the parameters behave very similarly in both emulators.

Despite this however, the parameter ranges are not nearly as well constrained as they are in \autoref{fig:Q19_SMF_results_2D}.
Furthermore, whereas the minimum implausibility subplots showed strong correlations before, they do not do this in \autoref{fig:Q19_full_results_SMF_2D}.
This is interesting, as this suggests that the SMF data and the LF/CMR data constrain \textit{different} regions of parameter space.
It however also implies that it would be rather challenging to perform a global parameter estimation on these parameters, as no region of parameter space is removed quickly.
Instead, there is a tiny hypercube shell of plausible samples that stretches over parameter space.

\begin{figure*}[htb!]
    \centering
    \subfloat{\includegraphics[width=0.24\textwidth]{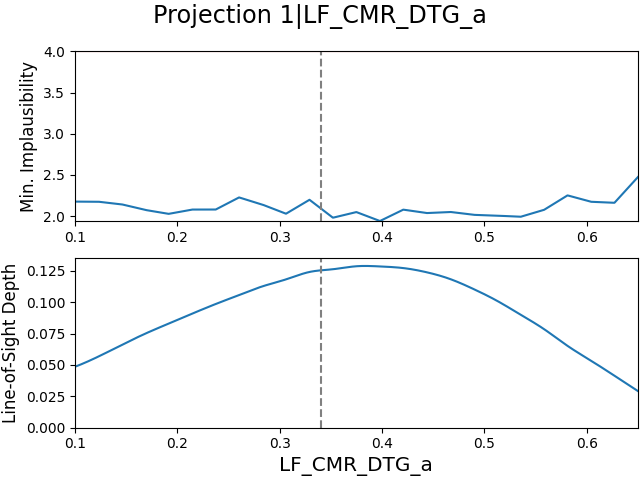}}
    \subfloat{\includegraphics[width=0.24\textwidth]{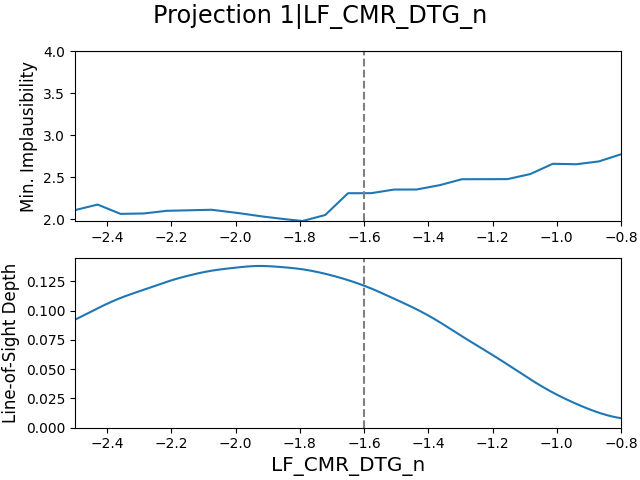}}
    \subfloat{\includegraphics[width=0.24\textwidth]{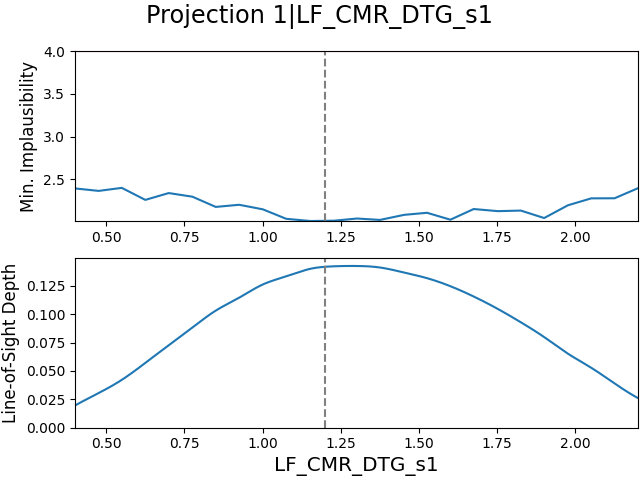}} \\
    \subfloat{\includegraphics[width=0.24\textwidth]{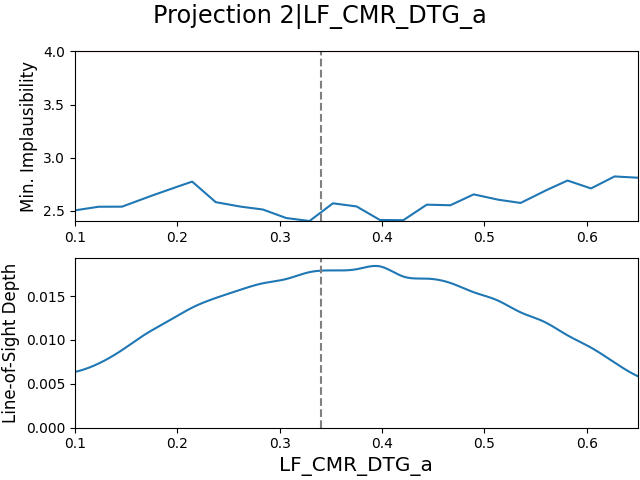}}
    \subfloat{\includegraphics[width=0.24\textwidth]{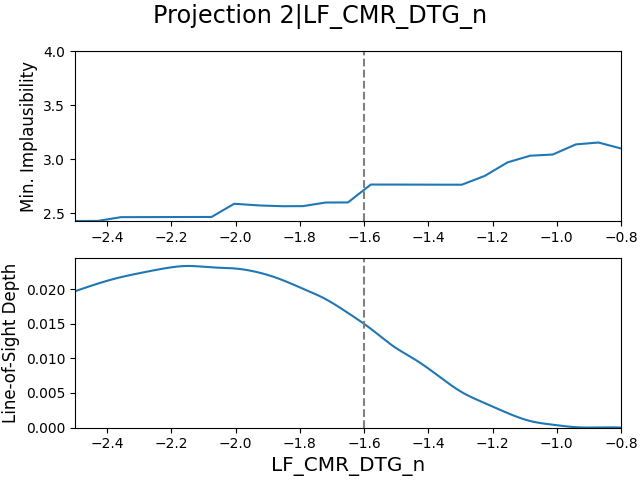}}
    \subfloat{\includegraphics[width=0.24\textwidth]{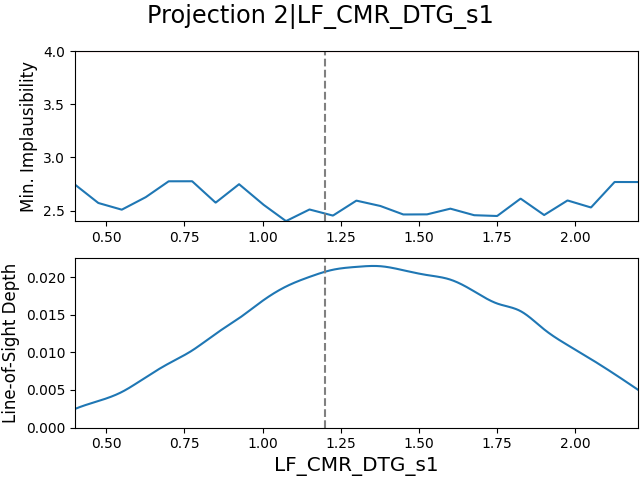}} \\
    \subfloat{\includegraphics[width=0.24\textwidth]{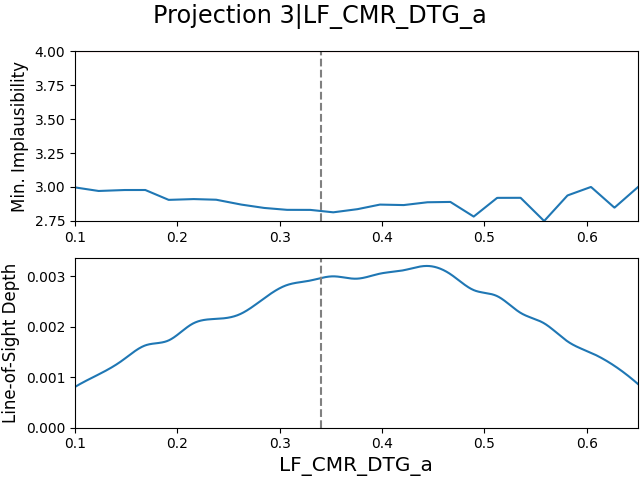}}
    \subfloat{\includegraphics[width=0.24\textwidth]{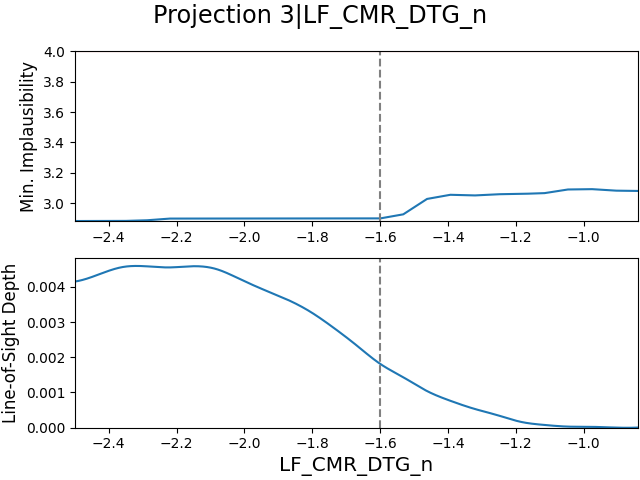}}
    \subfloat{\includegraphics[width=0.24\textwidth]{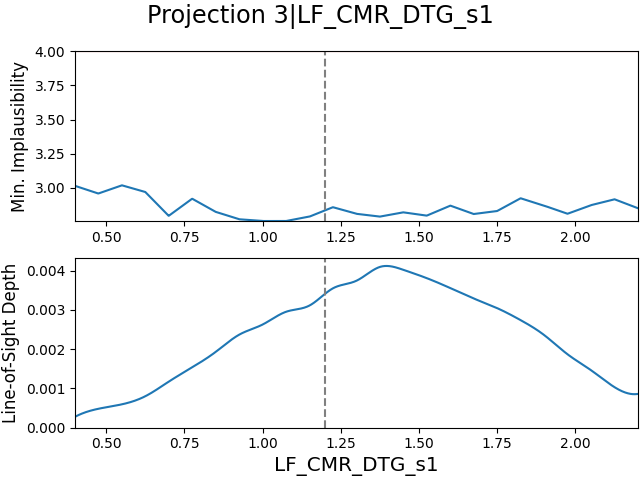}} \\
    \subfloat{\includegraphics[width=0.24\textwidth]{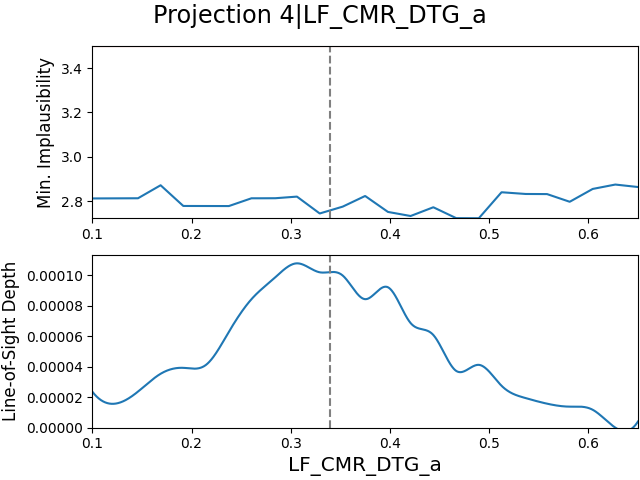}}
    \subfloat{\includegraphics[width=0.24\textwidth]{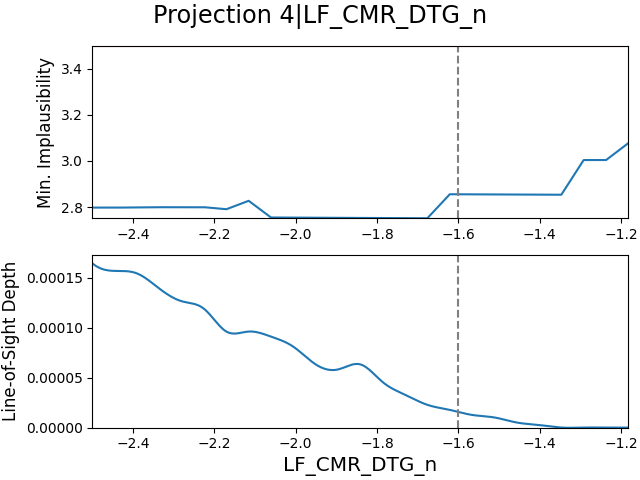}}
    \subfloat{\includegraphics[width=0.24\textwidth]{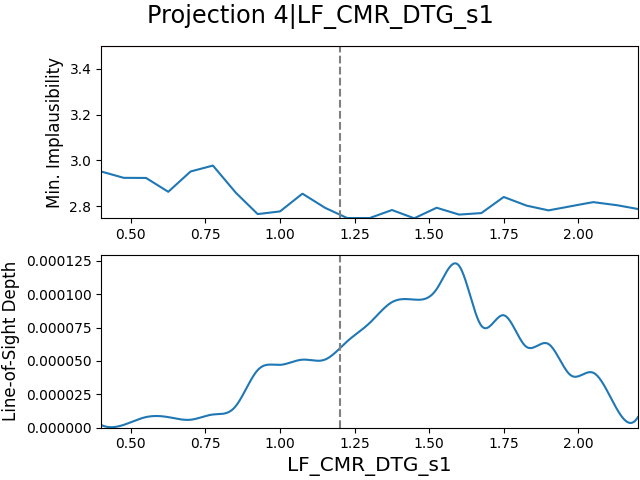}} \\
    \subfloat{\includegraphics[width=0.24\textwidth]{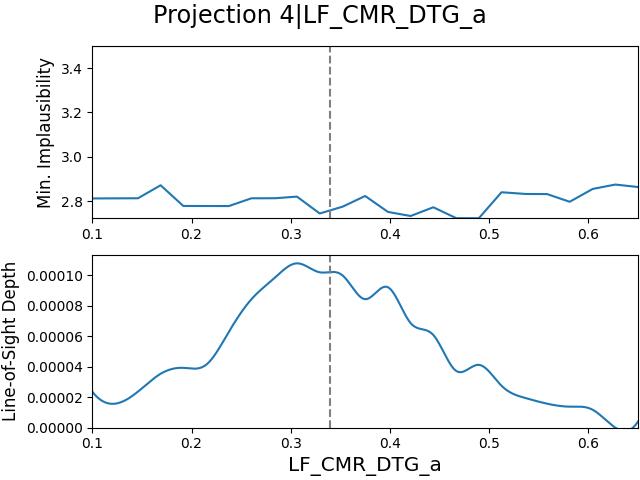}}
    \subfloat{\includegraphics[width=0.24\textwidth]{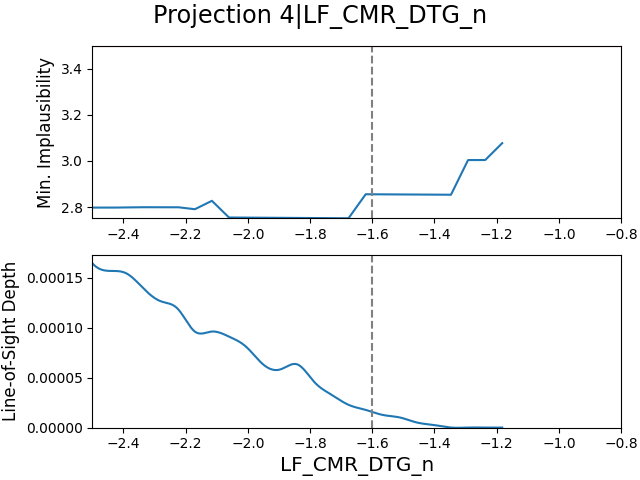}}
    \subfloat{\includegraphics[width=0.24\textwidth]{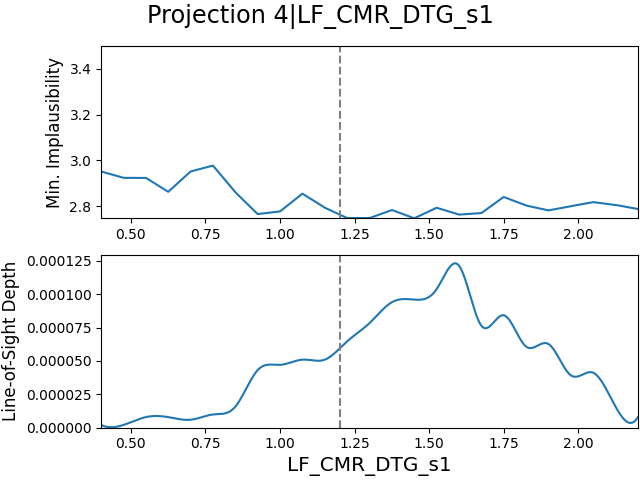}}
    \caption{2D projection figures of the full \citetalias{Qiu2019} \meraxes\ emulator at all four iterations, showing the three main free parameters related to the dust optical depth.
    The \textbf{dashed lines} show the estimated value of the corresponding parameter as given in \autoref{tab:Q19_par}.
    In case the parameter estimate is outside of the plotted value range, an \textbf{arrow} pointing in the direction of the estimate is shown instead.
    \textbf{First row:} $i_{\mathrm{emul}}=1$.
    \textbf{Second row:} $i_{\mathrm{emul}}=2$.
    \textbf{Third row:} $i_{\mathrm{emul}}=3$.
    \textbf{Bottom rows:} $i_{\mathrm{emul}}=4$ with the \textbf{final row} showing the full parameter range instead of only the defined range, but it is otherwise equivalent to the \textbf{fourth row}.}
    \label{fig:Q19_full_results_LF_2D}
\end{figure*}

In \autoref{fig:Q19_full_results_LF_2D}, we show the 2D projection figures of the three free parameters used for calculating the dust optical depth.
Note that the names of these parameters, as reported in the title of a projection figure, have an added \texttt{LF\_CMR\_DTG\_} prefix.
As these parameters are free parameters in a single equation, we do not expect them to show interesting details.

Looking at these projection figures, all the free parameters appear to agree with their estimates in all iterations, although \texttt{n} is slightly questionable.
However, with the exception of a small fraction for \texttt{n}, their plausible ranges are not reduced at all throughout the iterations.
As these parameters are free parameters, they are expected to be heavily correlated with each other and thus the behavior in the projection figures is logical.
It would however indicate a difficulty in finding their optimal values though.

\begin{figure*}[htb!]
    \centering
    \subfloat{\includegraphics[width=0.24\textwidth]{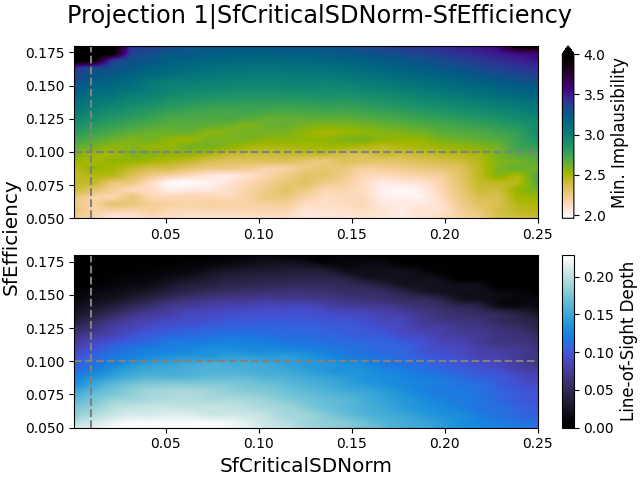}}
    \subfloat{\includegraphics[width=0.24\textwidth]{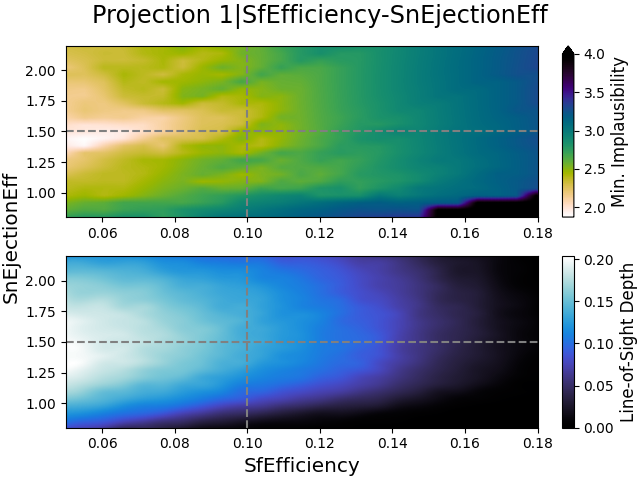}}
    \subfloat{\includegraphics[width=0.24\textwidth]{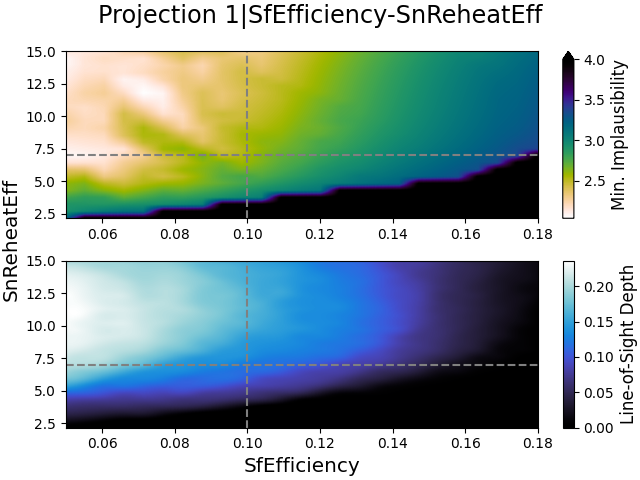}}
    \subfloat{\includegraphics[width=0.24\textwidth]{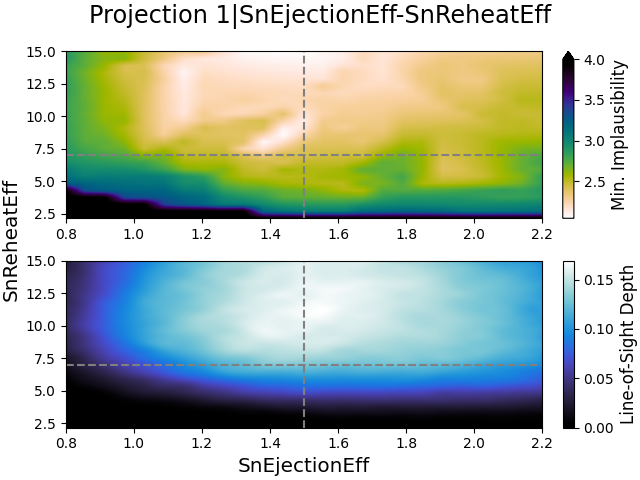}} \\
    \subfloat{\includegraphics[width=0.24\textwidth]{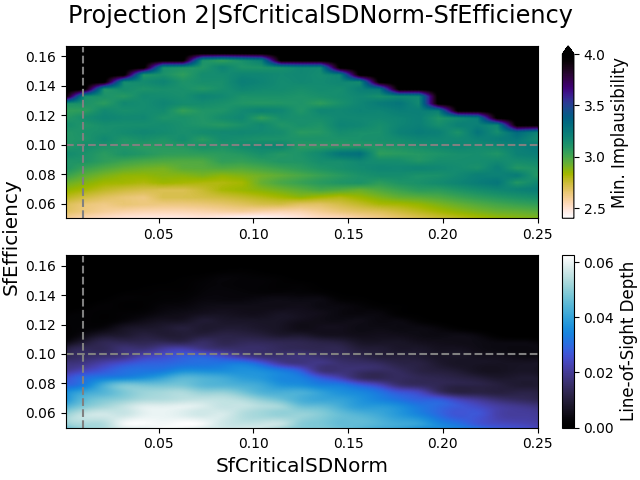}}
    \subfloat{\includegraphics[width=0.24\textwidth]{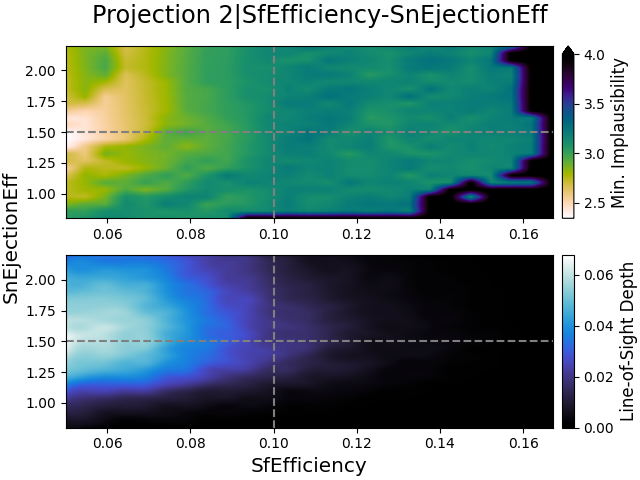}}
    \subfloat{\includegraphics[width=0.24\textwidth]{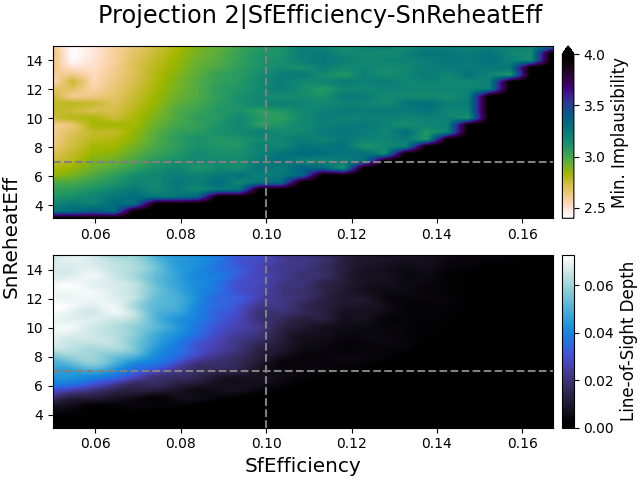}}
    \subfloat{\includegraphics[width=0.24\textwidth]{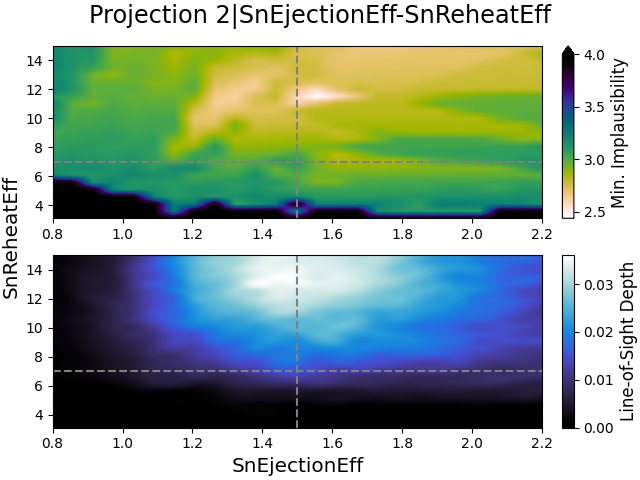}} \\
    \subfloat{\includegraphics[width=0.24\textwidth]{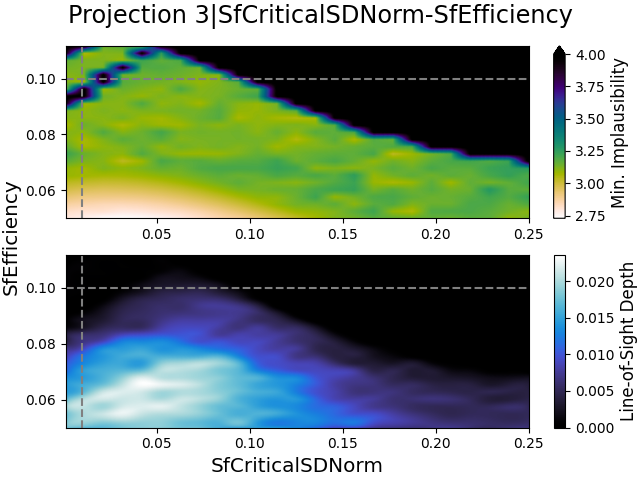}}
    \subfloat{\includegraphics[width=0.24\textwidth]{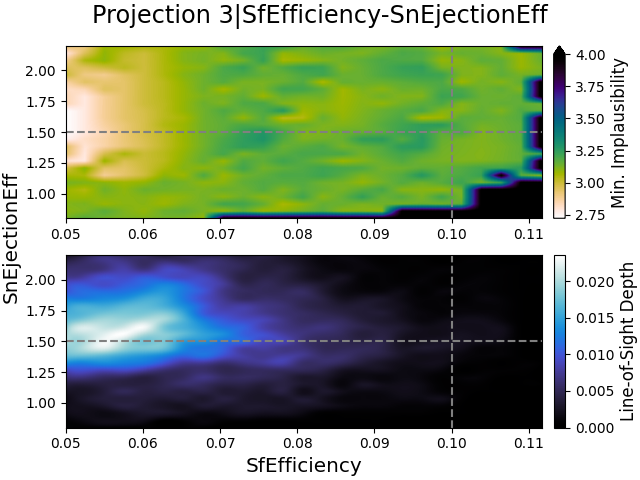}}
    \subfloat{\includegraphics[width=0.24\textwidth]{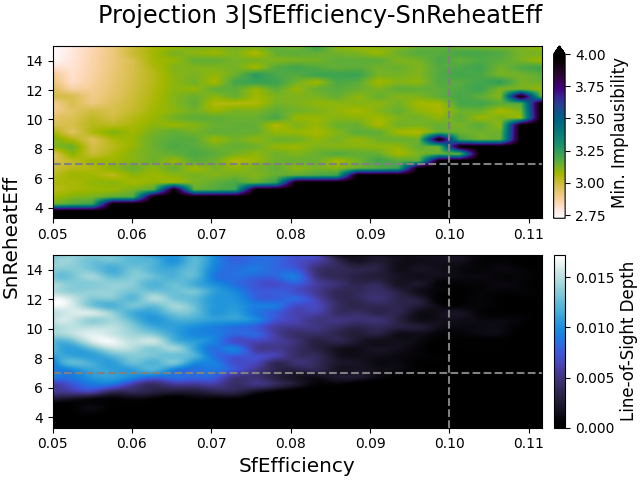}}
    \subfloat{\includegraphics[width=0.24\textwidth]{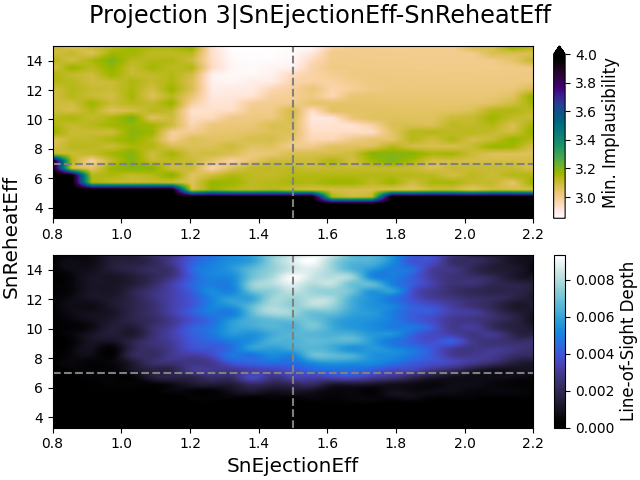}} \\
    \subfloat{\includegraphics[width=0.24\textwidth]{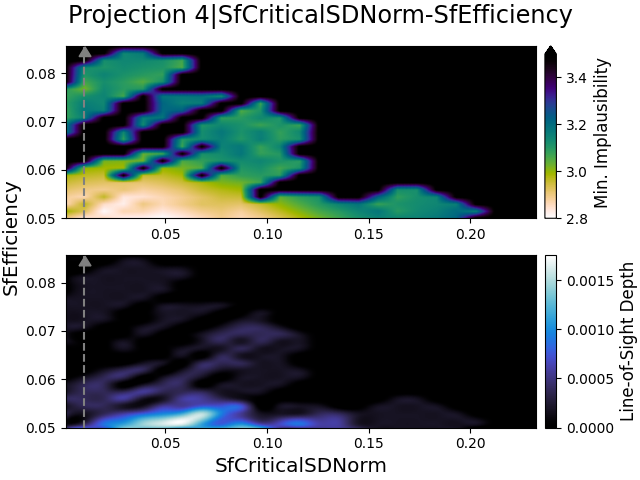}}
    \subfloat{\includegraphics[width=0.24\textwidth]{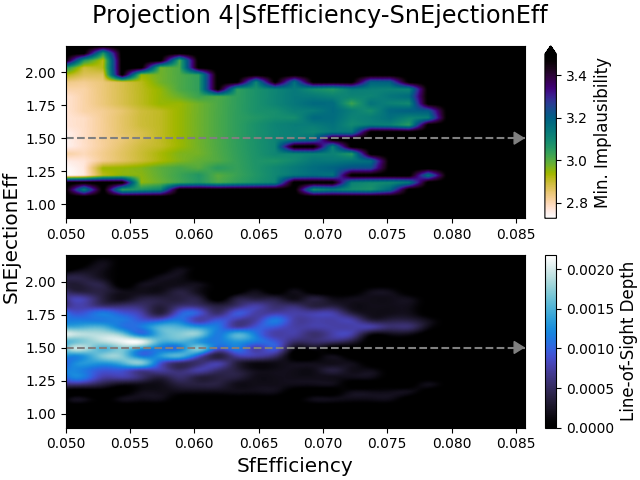}}
    \subfloat{\includegraphics[width=0.24\textwidth]{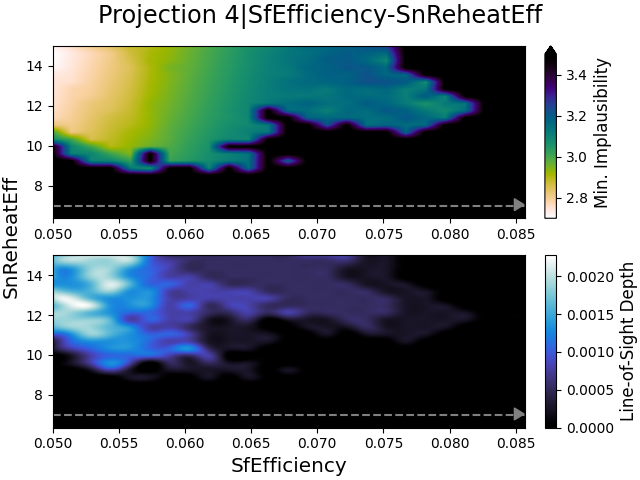}}
    \subfloat{\includegraphics[width=0.24\textwidth]{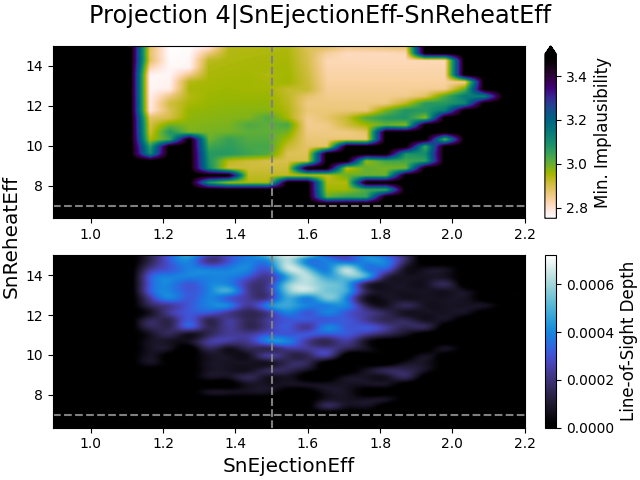}} \\
    \subfloat{\includegraphics[width=0.24\textwidth]{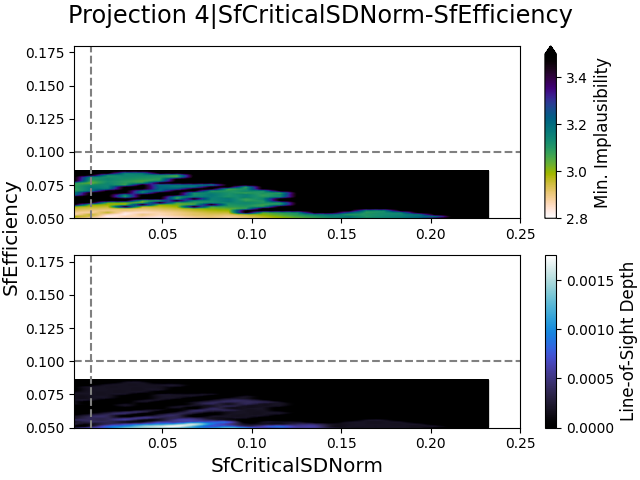}}
    \subfloat{\includegraphics[width=0.24\textwidth]{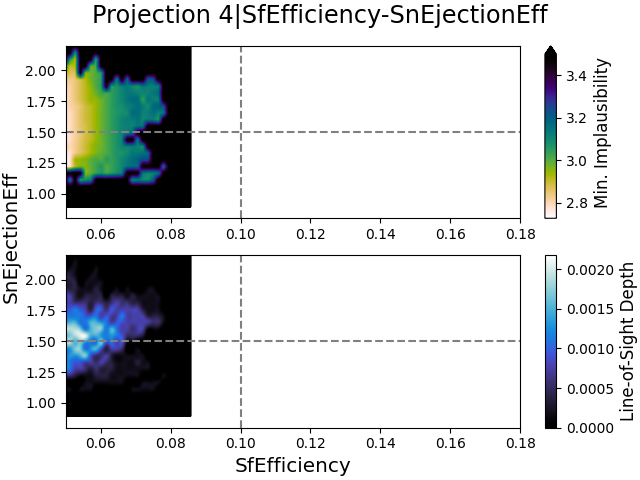}}
    \subfloat{\includegraphics[width=0.24\textwidth]{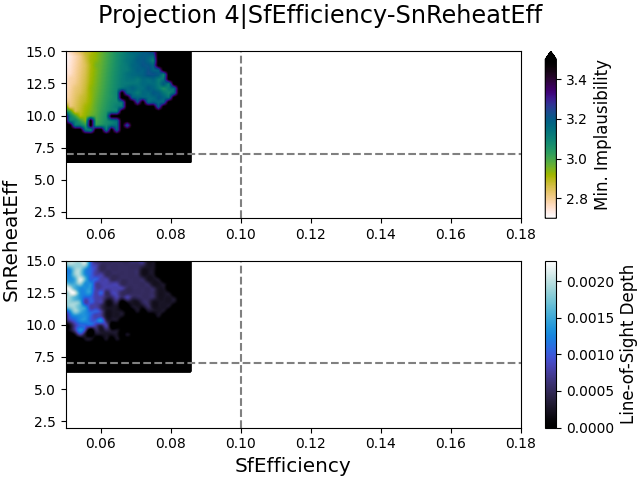}}
    \subfloat{\includegraphics[width=0.24\textwidth]{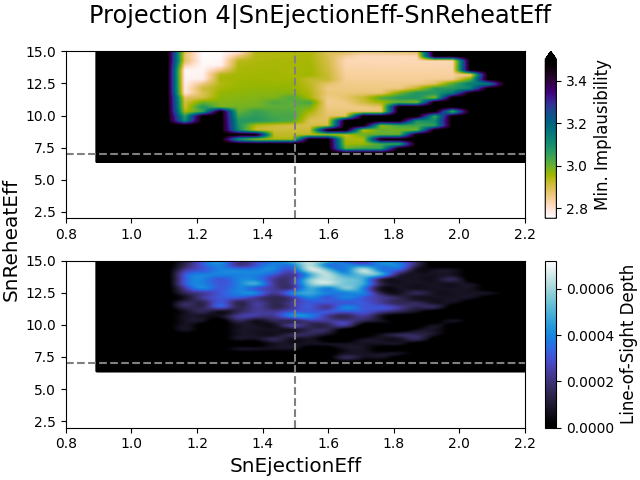}}
    \caption{3D projection figures of the full \citetalias{Qiu2019} \meraxes\ emulator at all four iterations, showing the four parameters related to the star formation and supernova feedback.
    The \textbf{first three columns} show the correlation between the star formation efficiency \texttt{SfEfficiency} and the other parameters.
    The \textbf{last column} shows the correlation between the energy coupling efficiency \texttt{SnEjectionEff} and the mass loading factor \texttt{SnReheatEff}.
    The \textbf{dashed lines} show the estimated value of the corresponding parameter as given in \autoref{tab:Q19_par}.
    In case either parameter estimate is outside of the plotted value range, an \textbf{arrow} pointing in the direction of the intersection of the estimates is shown instead.
    \textbf{First row:} $i_{\mathrm{emul}}=1$.
    \textbf{Second row:} $i_{\mathrm{emul}}=2$.
    \textbf{Third row:} $i_{\mathrm{emul}}=3$.
    \textbf{Bottom rows:} $i_{\mathrm{emul}}=4$ with the \textbf{final row} showing the full parameter range instead of only the defined range, but it is otherwise equivalent to the \textbf{fourth row}.}
    \label{fig:Q19_full_results_SMF_3D}
\end{figure*}

Similarly to the previous emulator, in \autoref{fig:Q19_full_results_SMF_3D}, we show the 3D projections for the parameters related to the star formation and supernova feedback processes, for all four emulator iterations.
As expected for the first emulator iteration, given its $f_{\mathrm{space}}$ value of $9.89\%$, the first row of projections shows nothing of great interest.
There are a few small indications of correlations between the parameters, but nothing significant.

\begin{figure*}[htb!]
    \centering
    \subfloat{\includegraphics[width=0.24\textwidth]{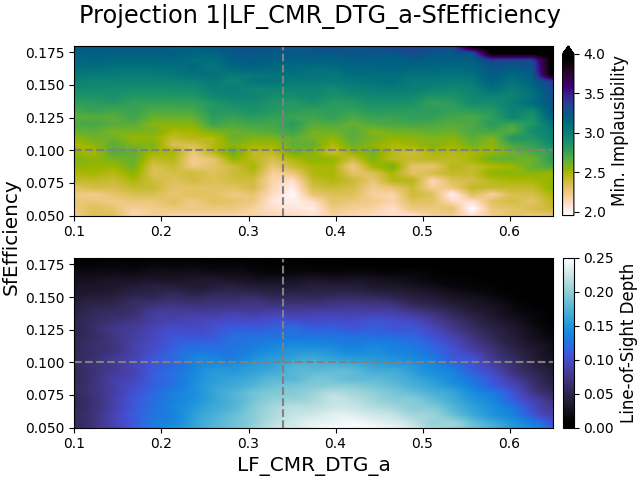}}
    \subfloat{\includegraphics[width=0.24\textwidth]{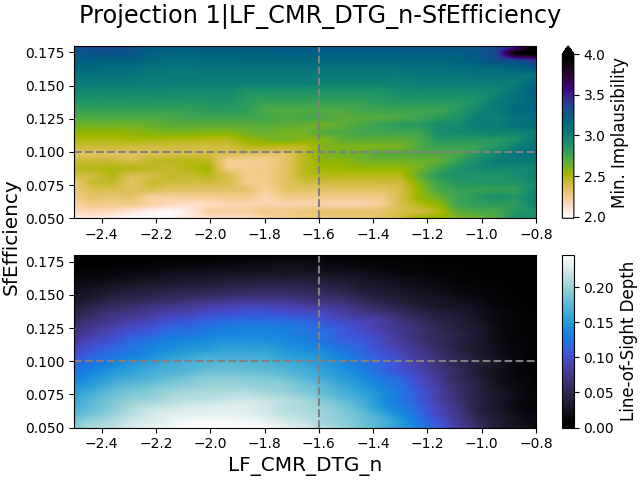}}
    \subfloat{\includegraphics[width=0.24\textwidth]{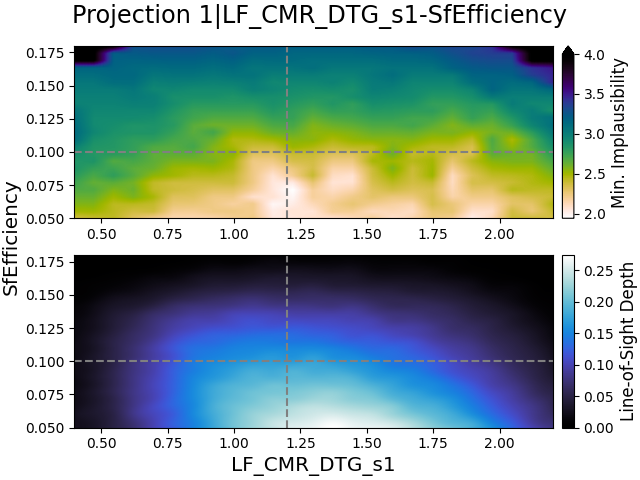}} \\
    \subfloat{\includegraphics[width=0.24\textwidth]{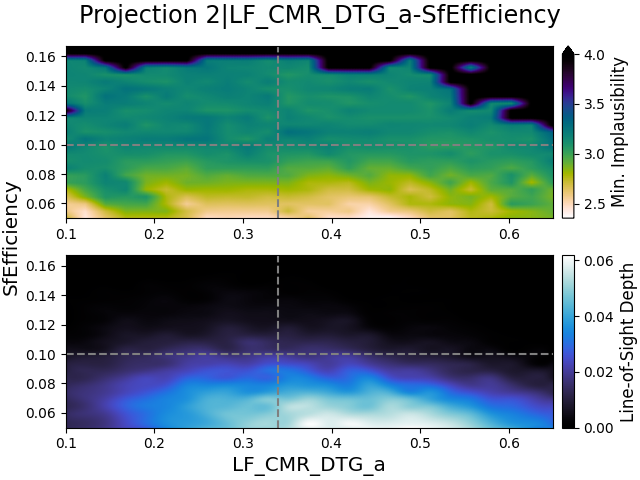}}
    \subfloat{\includegraphics[width=0.24\textwidth]{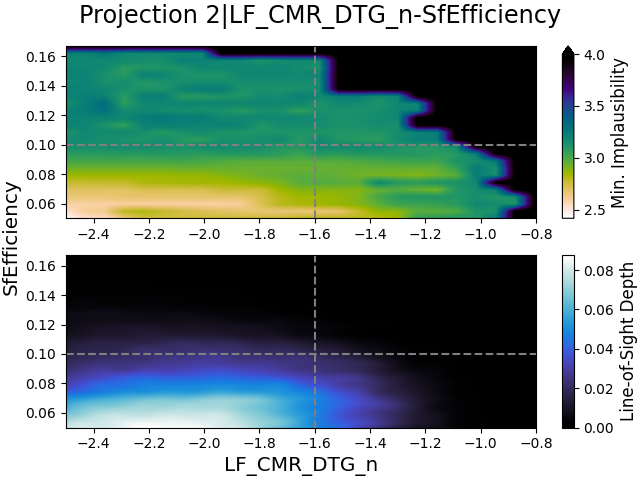}}
    \subfloat{\includegraphics[width=0.24\textwidth]{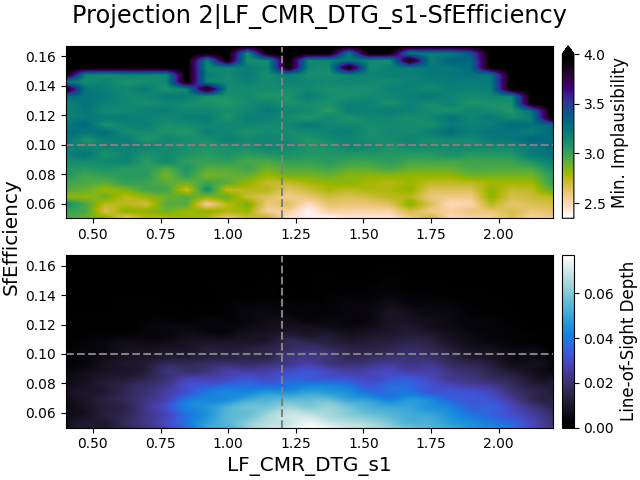}} \\
    \subfloat{\includegraphics[width=0.24\textwidth]{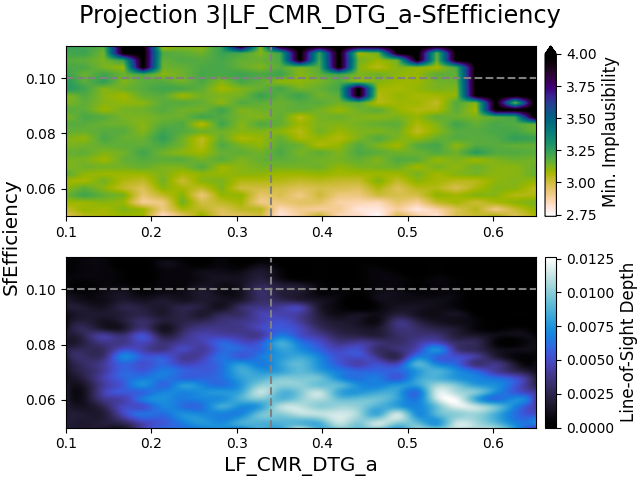}}
    \subfloat{\includegraphics[width=0.24\textwidth]{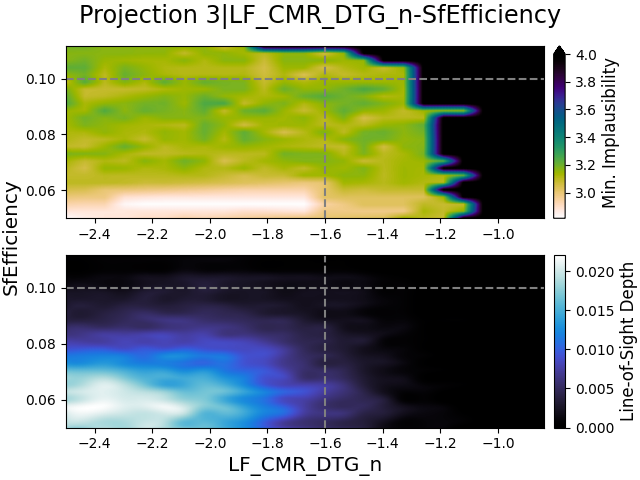}}
    \subfloat{\includegraphics[width=0.24\textwidth]{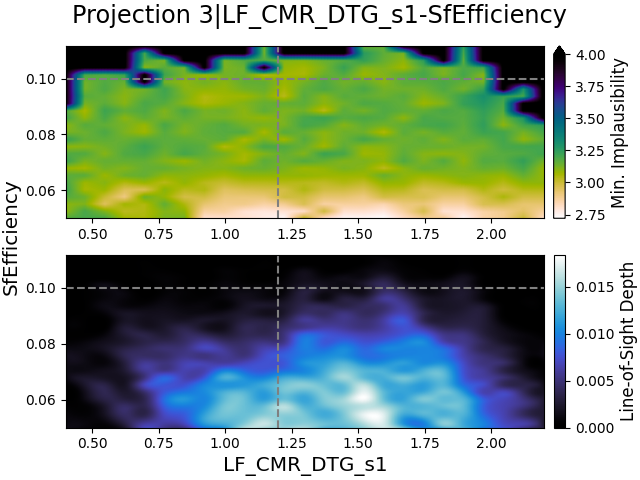}} \\
    \subfloat{\includegraphics[width=0.24\textwidth]{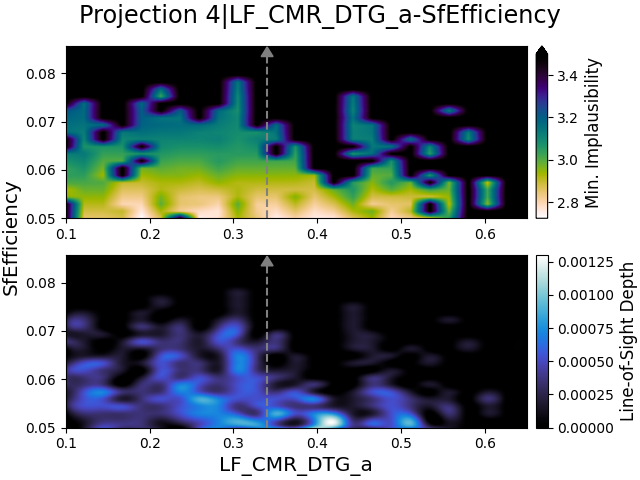}}
    \subfloat{\includegraphics[width=0.24\textwidth]{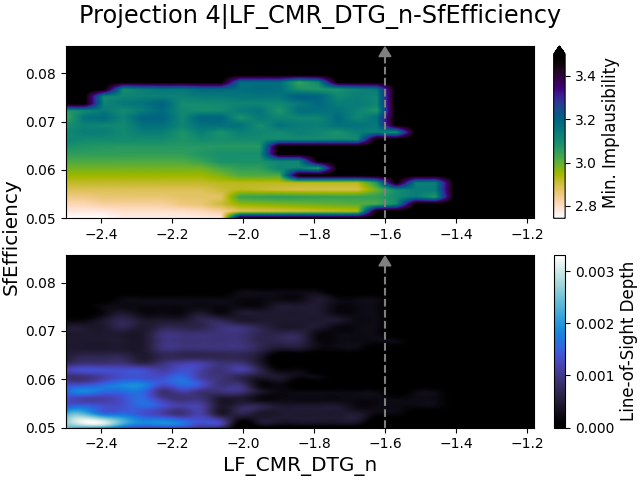}}
    \subfloat{\includegraphics[width=0.24\textwidth]{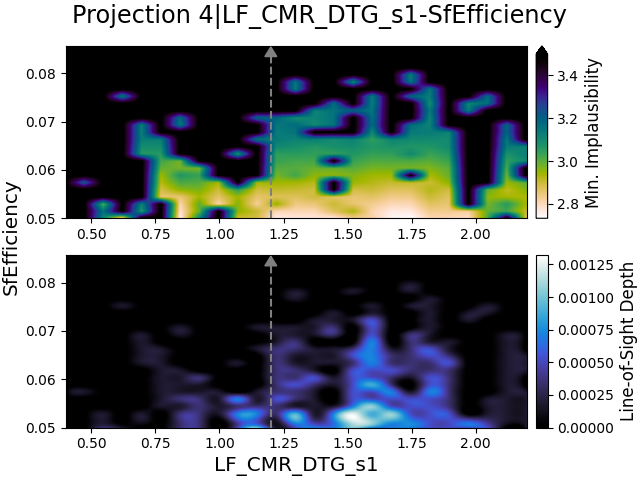}} \\
    \subfloat{\includegraphics[width=0.24\textwidth]{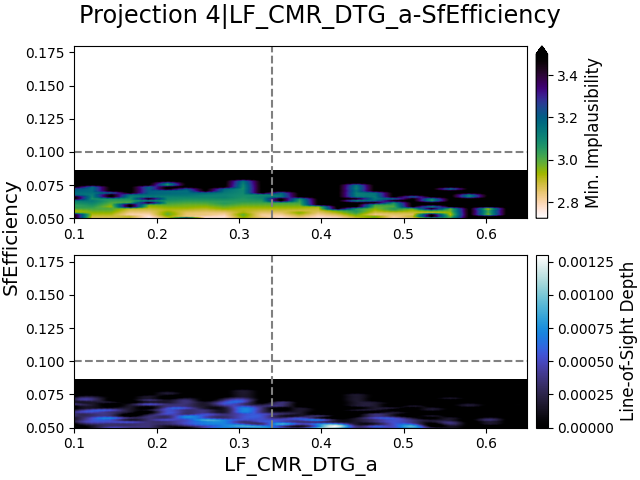}}
    \subfloat{\includegraphics[width=0.24\textwidth]{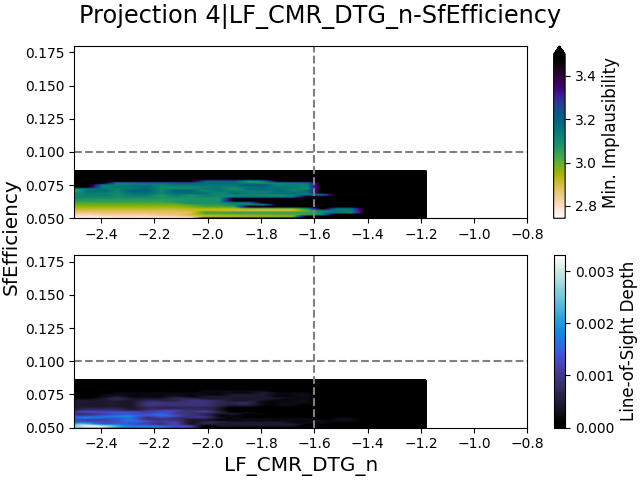}}
    \subfloat{\includegraphics[width=0.24\textwidth]{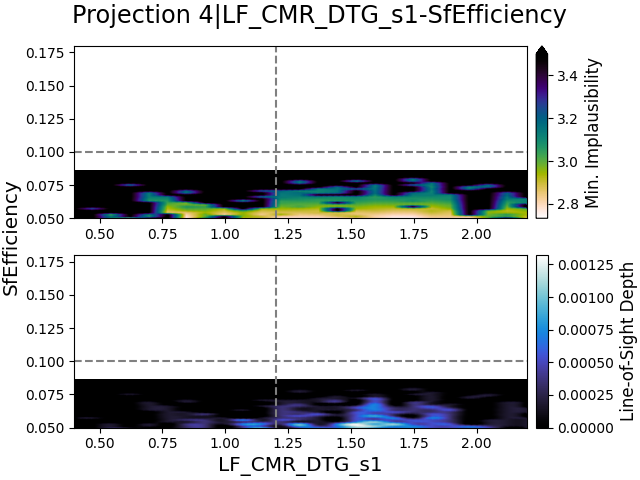}}
    \caption{3D projection figures of the full \citetalias{Qiu2019} \meraxes\ emulator at all four iterations, showing the three free parameters related to the dust optical depth.
    The \textbf{three columns} show the correlation between the star formation efficiency \texttt{SfEfficiency} and the main free dust optical depth parameters.
    The \textbf{dashed lines} show the estimated value of the corresponding parameter as given in \autoref{tab:Q19_par}.
    In case either parameter estimate is outside of the plotted value range, an \textbf{arrow} pointing in the direction of the intersection of the estimates is shown instead.
    \textbf{First row:} $i_{\mathrm{emul}}=1$.
    \textbf{Second row:} $i_{\mathrm{emul}}=2$.
    \textbf{Third row:} $i_{\mathrm{emul}}=3$.
    \textbf{Bottom rows:} $i_{\mathrm{emul}}=4$ with the \textbf{final row} showing the full parameter range instead of only the defined range, but it is otherwise equivalent to the \textbf{fourth row}.}
    \label{fig:Q19_full_results_LF_3D}
\end{figure*}

The second row of projection figures on the other hand, begins to exhibit some interesting behavior.
Here, we can see that there are regions in parameter space where better (i.e., lower minimum implausibility) samples can be found, but the plausible samples are still very spread out over parameter space.
The same can be observed for the third iteration.
In the last iteration however, there are actually significant parts of parameter space that are no longer considered to be plausible.
We also note that, unlike the projections in \autoref{fig:Q19_SMF_results_3D}, the star formation efficiency \texttt{SfEfficiency} no longer dominates the other parameters as strongly.
This further suggests that the SMF and LF/CMR data constrain different parts of parameter space, making it much harder to perform global parameter estimations.

And, finally, in \autoref{fig:Q19_full_results_LF_3D}, we show the 3D projection figures of the dust optical depth parameters.
Similarly to the 3D projections shown in \autoref{fig:Q19_full_results_SMF_3D}, it takes until iteration $4$ before parameter space finally begins to be restricted, despite only $0.00489\%$ of it being plausible.
Even then however, basically all values in the parameter ranges of the dust optical depth parameters are plausible, with no indication of converging.
Here, it appears that the observational data is not enough to constrain these free parameters.

\begin{figure*}[htb!]
    \centering
    \subfloat{\includegraphics[width=\textwidth]{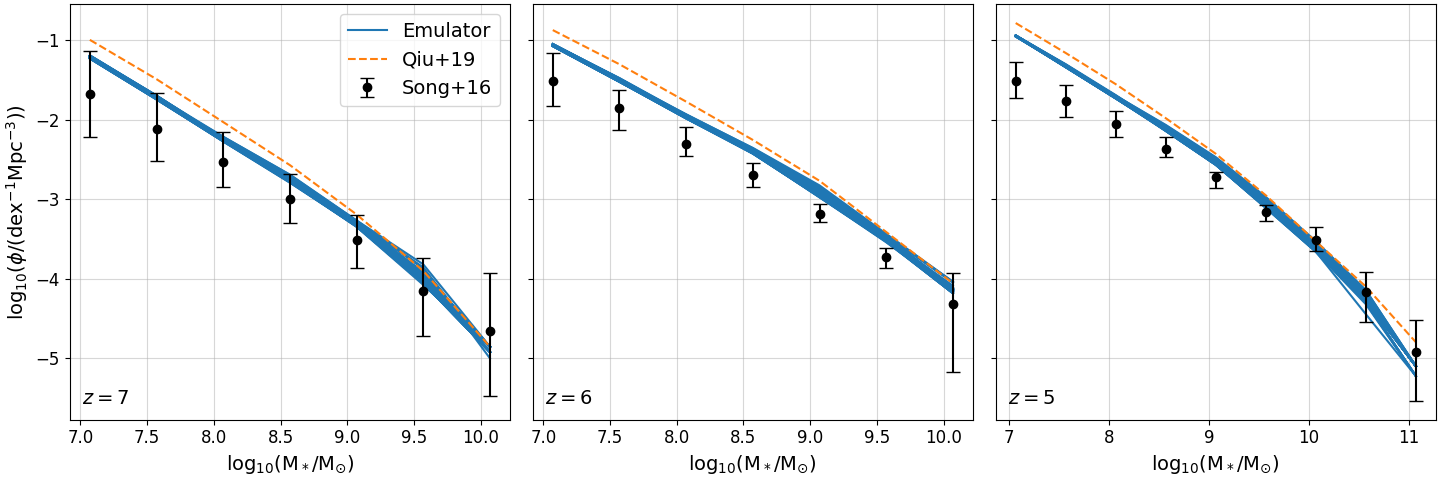}}\\
    \subfloat{\includegraphics[width=\textwidth]{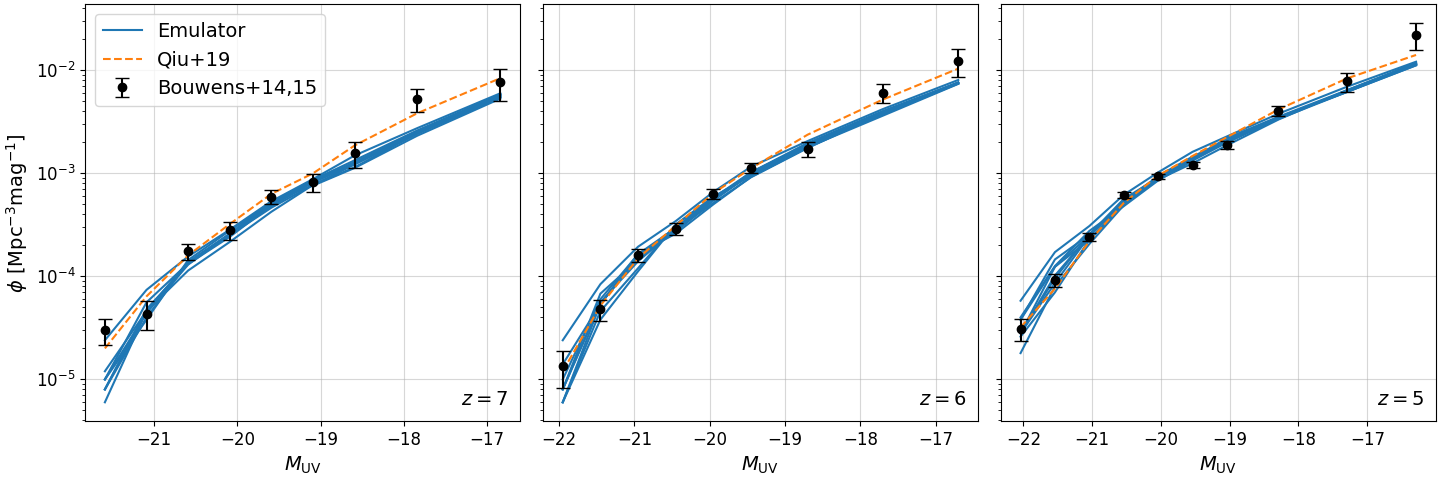}}\\
    \subfloat{\includegraphics[width=\textwidth]{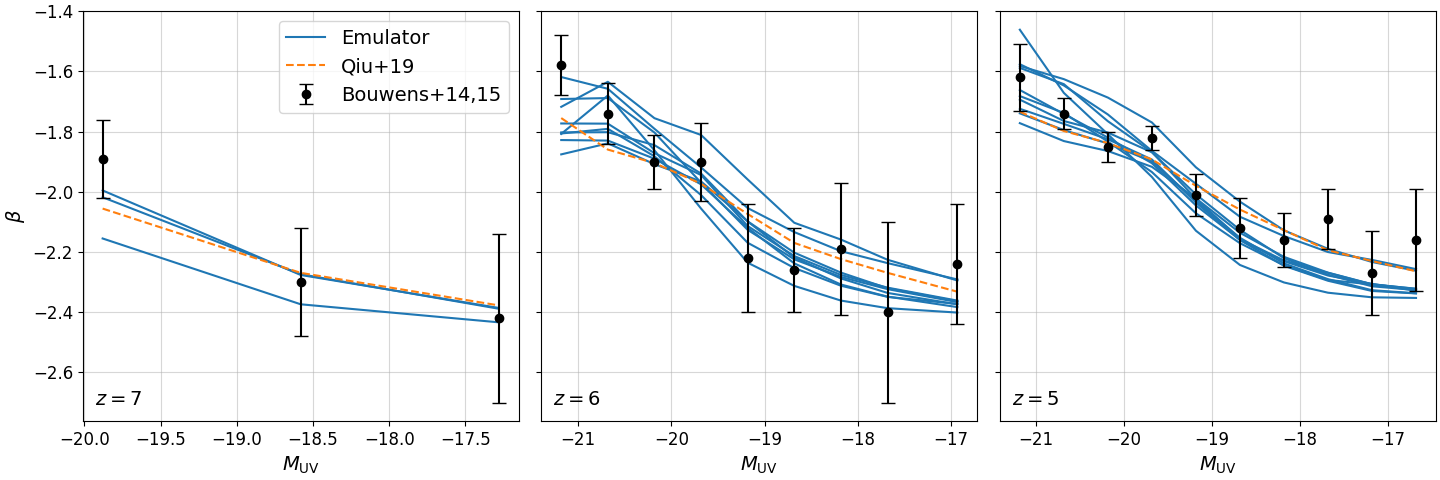}}
    \caption{Realizations of the stellar mass functions (\textbf{top}); luminosity functions (\textbf{center}) and color-magnitude relations (\textbf{bottom}) at redshifts $z=[7, 6, 5]$ using the results of the full \citetalias{Qiu2019} \meraxes\ emulator.
    All realizations were created by directly evaluating \citetalias{Qiu2019} \meraxes.
    The \textbf{solid lines} use the $50$ best plausible samples out of $2,500$ in the emulator at iteration $4$.
    The \textbf{dashed line} uses the parameter estimates as given in \autoref{tab:Q19_par}.
    The \textbf{dots} show the SMF data from \citet{Song2016} adjusted for a Kroupa IMF, or the LF/CMR data from \citet{Bouwens2014,Bouwens2015}, respectively, with corresponding standard deviations.}
    \label{fig:Q19_full_fits}
\end{figure*}

In \autoref{fig:Q19_full_fits}, we show a comparison between the SMFs/LFs/CMRs created by $50$ plausible samples in the emulator and the SMF/LF/CMR given by the parameter estimates from \autoref{tab:Q19_par}.
Similarly to \autoref{fig:Q19_SMF_fits}, we evaluated the emulator $2,500$ times within plausible space at iteration $4$, and selected the $50$ best plausible samples.
These $50$ samples were then evaluated in \citetalias{Qiu2019} \meraxes\ to produce the solid lines shown in the figure.

From this figure, we can see that the SMF fits are very similar (if not the same) as those we showed in \autoref{fig:Q19_SMF_fits}.
This is in agreement with the projection figures in \autoref{fig:Q19_SMF_results_2D} and \autoref{fig:Q19_full_results_SMF_2D}, which also showed similar parameter trends.
However, whereas the emulator samples provide better fits for the SMF, it appears to provide roughly equal or slightly worse fits for the LF (the center row in \autoref{fig:Q19_full_fits}), while the CMR fits seem to be equal (bottom row).
This implies that the SMF data constrains the LF much more heavily than the LF data itself does, but also that \citetalias{Qiu2019} \meraxes\ might be unable to fit both the SMF data and the LF/CMR data simultaneously.

When we take into account the information shown in all projection figures and the two realization figures, we come to the conclusion that the prior parameter ranges as given in \autoref{tab:Q19_par} are too restrictive; and that the SMF and LF/CMR data constrain the \meraxes\ model differently.
The former is clearly shown in the star formation and supernova feedback parameter projections in \autoref{fig:Q19_SMF_results_2D} and \autoref{fig:Q19_full_results_SMF_2D}.
In addition to the energy coupling efficiency, \texttt{SnEjectionEff}, all parameters have their best values at or very near the boundaries of their priors for both emulators.
The projection figures in \autoref{fig:Q19_full_results_SMF_2D} and \autoref{fig:Q19_full_results_LF_2D} on the other hand, show that the parameters are much harder to constrain when the LF/CMR data is used as well.

These findings demonstrate two significant results: The importance of having the correct data points when performing parameter estimations; and the value of using \prism\ for exploring models.
Since only the LF/CMR data from \citet{Bouwens2014,Bouwens2015} was used by \citetalias{Qiu2019} to constrain all nine parameters, the differing, stronger constraining power of the SMF data was not noted.
As we can see in the 2D projections in \autoref{fig:Q19_full_results_SMF_2D}, the best values for the star formation and supernova feedback parameters are similar to those in \autoref{fig:Q19_SMF_results_2D}, implying that the SMF data constrains these parameters more heavily than the LF/CMR data.
The two emulators discussed in this section required roughly $7,000$ evaluations of the \meraxes\ model, multiple orders of magnitude less than the average amount used by an MCMC approach.
Despite this however, the emulators still show us all this crucial information. 

In addition to the importance of using the correct data points, the projections in \autoref{fig:Q19_full_results_SMF_3D} highlight the potential dangers of having too many degrees-of-freedom in a model (or, alternatively, the dangers of not having enough constraining data points).
Even though the SMF and the LF/CMR data constrain the \meraxes\ model differently, causing a wide range of parameter values to be plausible, we can see much more structure in these projections than in those we showed in \autoref{fig:M16_SMF_results}.
We therefore also note the importance of choosing the degrees-of-freedom for a scientific model wisely.

\section{Conclusions}
\label{sec:Conclusions}
We have analyzed the galaxy formation model \meraxes\ using the emulation-based framework \prism.
Two different versions of \meraxes\ have been studied: the original created by \citetalias{Meraxes} and a modified one by \citetalias{Qiu2019}.
Both versions have provided us with different, interesting results.

The \citetalias{Meraxes} \meraxes\ model has shown to be difficult to constrain properly using the SMF observational data at $z = [7, 6, 5]$ alone.
As described in \ref{sec:M16}, there are several indications that imply that the \citetalias{Meraxes} \meraxes\ model has too many free parameters, mainly for the supernova feedback parameters \texttt{SnEjection\dots} (supernova energy coupling efficiency) and \texttt{SnReheat\dots} (mass loading factor) to be able to explain these SMF observations.
We also come to the conclusion that the manually calibrated parameter values for \citetalias{Meraxes} \meraxes\ are likely to be biased as the parameters do not converge, which is probably caused by the large correlations between the free parameters.
\prism\ has demonstrated that it can quickly perform a rough parameter estimation using only a few thousand model evaluations, which provided us with the mentioned results.

In \ref{sec:Q19}, we analyzed the \citetalias{Qiu2019} \meraxes\ model, which has a reduced parameter space for the star formation parameters, and uses only a single degree-of-freedom for both supernova feedback parameters.
Our analysis has shown that this greatly improves the convergence rate of the \meraxes\ model and that it greatly facilitates constraining \meraxes, as shown in \autoref{tab:Q19_SMF_stats} and \autoref{fig:Q19_SMF_results_3D}.
This figure also showed us the value of using the performance of acceptable parameter values (e.g., minimum implausibility) in addition to their density when performing parameter estimations.

Afterward, in \ref{subsec:Q19_full}, we constrained the \citetalias{Qiu2019} \meraxes\ model using both SMF and LF/CMR data.
This gave us two important results: The SMF and the LF/CMR data constrain the \meraxes\ model differently; and using heavily constrained parameter priors can bias the best parameter fits.
The latter was shown in \autoref{fig:Q19_SMF_results_2D} and \autoref{fig:Q19_full_results_SMF_2D}, where the best parameter values can be found very near or at the boundary of their priors.
Alternatively, when looking at this from a physics standpoint, this can imply that aspects of the \meraxes\ model itself are incomplete or incorrect if the used priors are determined by their physical meaning, which is a rather important detail to know.

The results in \ref{sec:M16} and \ref{sec:Q19} show us the importance of using more than one observational data type when constraining models, especially when appropriate observational data is scarce.
Additionally, the choice of the degrees-of-freedom in a model should be made wisely as well.
With our results, we have demonstrated the potential of using an emulation-based framework like \prism\ for analyzing scientific models, which is a task that a full Bayesian analysis cannot perform quickly.
The use of \prism\ allows us to discover important details such as those mentioned above at an early stage and we therefore recommend it as a core component in the development of scientific models.

\section*{Acknowledgements}
E.v.d.V.\ would like to thank Michael Goldstein, Manodeep Sinha and Ian Vernon for fruitful discussions and valuable suggestions.
Parts of the results in this work make use of the \textit{rainforest} and \textit{freeze} colormaps in the \textsw{CMasher} package \citep{cmasher}.
We are thankful for the open-source software packages used extensively in this work, including \textsw{e13tools}\footnote{\url{https://github.com/1313e/e13Tools}}; \textsw{h5py} \citep{h5py}; \textsw{hickle} \citep{hickle}; \textsw{matplotlib} \citep{matplotlib}; \textsw{mlxtend} \citep{mlxtend}; \textsw{mpi4py} \citep{mpi4py}; \textsw{mpi4pyd}\footnote{\url{https://github.com/1313e/mpi4pyd}}; \textsw{numpy} \citep{numpy}; \textsw{scikit-learn} \citep{scikit-learn}; \textsw{scipy} \citep{scipy}; \textsw{sortedcontainers} \citep{sortedcontainers}; and \textsw{tqdm}\footnote{\url{https://github.com/tqdm/tqdm}}.
Parts of this research were supported by the Australian Research Council Centre of Excellence for All Sky Astrophysics in 3 Dimensions (ASTRO 3D), through project number CE170100013.
Parts of this work were performed on the OzSTAR national facility at Swinburne University of Technology. OzSTAR is funded by Swinburne University of Technology and the National Collaborative Research Infrastructure Strategy (NCRIS).



\DeclareRobustCommand{\DUTCH}[3]{#3}    
\bibliographystyle{aasjournal}
\bibliography{bibliography}


\end{document}